\documentclass[iop]{emulateapj}  

\usepackage{graphicx, natbib, color, bm, amsmath, epsfig,ctable}

\usepackage{graphicx, natbib, color, bm, amsmath, epsfig}

\newcommand{\avir}{\ensuremath{\alpha_{\rm{vir}}}}

\newcommand{\msun}{\ensuremath{M_\odot}}

\def\rms{\ensuremath{\rm{rms}}}

\newcommand{\tff}{\ensuremath{t_{\rm{ff}}}}

\definecolor{orange}{rgb}{1.        ,  0.54,  0}

\definecolor{meta}{rgb}{0.371,0.617,0.625} 

\newcommand{\dc}[1]{}
\def\Bvec{\ensuremath{{\bf B}}}

\def\vvec{\ensuremath{{\bf v}}}

\def\kvec{\ensuremath{{\bf k}}}

\definecolor{pink}{rgb}{1.        ,  0.75294118,  0.79607843}
\definecolor{maroon}{rgb}{0.69019608,  0.18823529,  0.37647059}

\definecolor{gray}{rgb}{0.5,0.5,0.5}

\newcommand{\percc}{\ensuremath{\rm{cm}^{-3}}}

\newcommand{\kms}{\ensuremath{\rm{km}\ \rm{s}^{-1}}}

\def\sa{super-Alfv\' enic}

\def\cs{\ensuremath{c_{\rm{s}}}}

\def\hw{0.49}

\def\betao{\ensuremath{\beta_0}}

\renewcommand\vvec{\ensuremath{ {\bf v}}}

\newcommand{\hltwo}[1]{#1}

\begin{document}
\title{Observational Diagnostics of Self-Gravitating MHD Turbulence in Giant Molecular Clouds}
\author{Blakesley Burkhart \altaffilmark{1}, David C. Collins \altaffilmark{2}, 
Alex Lazarian \altaffilmark{3}}

\altaffiltext{1}{Harvard-Smithsonian Center for Astrophysics, 60 Garden St., Cambridge, MA 0213}
\altaffiltext{2}{Department of Physics, Florida State University, Tallahassee,
FL 32306-4350}
\altaffiltext{3}{Astronomy Department, University of Wisconsin, Madison, 475 N.
Charter St., WI 53706, USA}
\begin{abstract}
We study the observable signatures of self-gravitating MHD turbulence 
by applying the probability density functions (PDFs) and the spatial density power spectrum to synthetic column density maps.
We find that there exists three characterizable stages of the evolution 
of the collapsing cloud which we term ``early," ``intermediate," and ``advanced."
At early times, i.e. $t<0.15t_{ff}$,
the column density has a power spectral slope similar to nongravitating supersonic turbulence and a lognormal distribution. 
At an intermediate stage, i.e. $0.15t_{ff}< t \leq 0.35t_{ff}$, 
there exists signatures of the prestellar cores
in the shallower PDF and power spectrum power law slopes.
The column density PDF power law tails at these times have line of sight averaged slopes ranging from -2.5 to -1.5
with shallower values belonging to simulations with lower magnetic field strength.
The density power spectrum slope becomes shallow and can be characterized by 
$P(k)=A_1k^{\beta_2}e^{-k/k_c}$, where $A_1$ describes the amplitude,
$k^{\beta_2}$ describes the classical power law behavior and the scale $k_c$
characterizes the turn over from turbulence dominated to self-gravity dominated.
At advanced stages of collapse, i.e. $\approx t>0.35t_{ff}$, the  power spectral slope is positive valued,
and a dramatic increase is observed in the PDF moments and the Tsallis incremental PDF parameters,
which gives rise to deviations between PDF-sonic Mach number relations.
Finally, we show that the imprint of gravity on the density power spectrum 
can be replicated in non-gravitating turbulence by 
introducing a delta-function with amplitude equivalent to the maximum valued point in a given self-gravitating map.
We find that the turbulence power spectrum restored through spatial filtering of the high density material.  
\end{abstract}

\maketitle

\section{Introduction}
Molecular clouds are highly turbulent, magnetized and are the sites of all known star formation \citep{Elmegreen04}.
The details of
the collapse of molecular clouds determine the key properties of the star
formation rate (SFR) and stellar initial mass distribution 
\citep[IMF, see e.g.][]{Hennebelle12}
Thus, the development of a detailed understanding of the dynamics of molecular clouds is
an essential step toward a complete picture of star formation, including predicting the initial mass function.

The turbulent nature of molecular clouds is evident from a variety of observations including
non-thermal broadening of molecular emission and absorption lines such as carbon monoxide 
\citep[see][]{Spitzer78, Stutzki90, Heyer04}
and fractal and hierarchical structures 
\citep[see][]{Elmegreen83, Vazquez-Semadeni94, Burkhart13}
A number of
new techniques, including those studying the turbulence velocity spectrum
\citep[see][for a review]{Lazarian09} and the sonic Mach number
and Alfv\'en Mach number \citep[see][]{Kowal07, Burkhart09, Burkhart10,
Burkhart12b,  Esquivel10, 
Tofflemire11}, have been applied to Giant Molecular Clouds (GMCs) and also have shown that turbulence there is supersonic.
In light of this, turbulence seems to play a duel role in the GMC environment of providing support
on the large scales (i.e. scales of the cloud) while compressing small scales via shocks \citep{MacLow04}.
Although it is clear that molecular clouds are turbulent \citep{Elmegreen04,MacLow04}, magnetized
\citep{Crutcher12} and self-gravitating \citep{Larson81,Solomon87}, 
the relative importance of these components is still under debate \citep[e.g.][]{Padoan99,Li09,Vazquez08, Kritsuk13,
Li14b}.  As a mature theory of magnetized, supersonic, self-gravitating turbulence has
yet to emerge, interpretations of observations are difficult.
In light of this, numerical models have proven to be an important tool in the understanding of
observations.  

Several recent numerical studies have probed the observational signatures of
turbulent clouds, both with and without self-gravity and with a range of Mach numbers
and magnetic field strengths.
Simulations, observations and theoretic works have all pointed out the fact that
compressible turbulence is important for creating filaments and regions of high density contrast
\citep{Kowal07, Burkhart09, Federrath10}.  Shocks can broaden the density/column density probability
density function (PDF) and shallow the power spectral slope.
The column density PDF and power spectrum (or
delta variance, which is isomorphic to the power spectrum) are the chief targets for
this study, due to their importance to turbulence based star formation theories
\citep{Padoan95, Krumholz05, Federrath13b}.  Both the  density PDF and
power spectrum have been shown to be sensitive to Mach number, magnetic field
strength, and self gravity \citep{Beresnyak05, Kowal07, Kainulainen09, Burkhart10, Burkhart12, Collins12,
Federrath13} however less attention has been paid to higher order statistics of column density. 
 
In this work we  explore the variation of the
properties of the PDFs and the power spectrum of column density as simulated clouds collapse under their own self-gravity as a function of 
time and magnetic field strength. The goal of this paper is to provide observers
with useful methods to apply to dust extinction maps and integrated intensity
maps,
and to explore if the PDFs/spectrum of column density behave similarly to 3D
density studied in previous works such as \citet{Collins12}.
In particular we are interested to explore if the collapse evolution of the cloud can be observationally discerned and pay particular attention to the time 
dependency of the turbulence statistics.
Further, we explore two novel fits to the
column density power spectrum, and use one of these techniques to disentangle
the signatures of self-gravity from the turbulence and to shed light on the origin of the increase in the  density power spectrum slope
\citep{Federrath13}.
The ultimate aim of this paper is to provide
set of tools to quantify the physical state of observed molecular clouds.

The layout of the paper is as follows.  In Section \ref{sec.method} we describe
the simulations studied and the codes used to generate them.  In Section
\ref{sec.columndensity} we introduce the the column density PDF and apply it to our simulations. 
We investigate the  evolution of the PDF
power law tail (Section \ref{sec.powerlaw}), the evolution of the variance
(Section \ref{sec.variance}), higher order moments (Section
\ref{sec.moments}) and finally the Tsallis distribution (Section
\ref{sec.tsallis}).  In Section \ref{sec.spectra} we discuss the evolution of
the slope of the power spectrum, a new fit to the evolving spectrum (Section \ref{sec.fit}); a technique to reproduce the self-gravitating spectrum from
the turbulent spectrum (Section \ref{sec.SGtoNSG}); and a technique to recover
the turbulent spectrum from the self-gravitating spectrum (Section
\ref{NSG_FROM_SG}).  We discuss the implications of our results in  Section \ref{sec.discussion} followed by our conclusions
in Section \ref{sec.conclusions}.

\section{Method}
\label{sec.method}

Two suites of simulations were used for this paper: one suite of non-gravitating
simulations that used a fixed resolution of $512^3$ (which we refer to as the Godunov simulations), and one suite of
self-gravitating simulations that used a $512^3$ root grid and four levels
of adaptive mesh refinement (AMR, which we refer to as the Enzo simulations).  All simulations solved the ideal MHD
equations with large-scale solenoidal forcing.

The Godunov simulations used here are
ten non-gravitating simulations with sonic
Mach numbers $\approx 0.5-22$ and have been used in many previous works
\citep{Cho03, Burkhart09, Burkhart10,
Kowal07, Kowal09, Kowal11}.  These simulations used the algorithm described in \citep{Cho02}, which uses a combination of spatially
third-order essentially nonoscillatory (ENO) methods to ensure both accuracy and
stability, and a third-order Runga-Kutta time integration. 
The Godunov models are divided into two groups
corresponding to sub-Alfv\'enic ($B_\mathrm{ext}=1.0$) and super-Alfv\'enic
($B_\mathrm{ext}=0.1$) turbulence. 

The three self-gravitating simulations were performed using the 
constrained transport MHD option in Enzo (MHDCT) \citep{Collins10, Enzo13}.  This
code uses the second order hyperbolic solver described by \citet{Li08a}, the
CT method of \citet{Gardiner05}, and
the divergence free interpolation method of \citet{Balsara01}.  These
simulations were described in detail in \citet{Collins12}.
The self-gravitating Enzo models are 
supersonic and \sa, with mass chosen such that kinetic and gravitational
energies are equal.  They have three different values of initial magnetic field which sets the Alfv\'en Mach number.  
Throughout the text and in Table 1, we denote these different cases as  low,
mid, and high to stand for ``low-valued," ``middle-valued," and ``high-valued" magnetic field runs.

\begin{figure*} \begin{center}
\includegraphics[width=0.49\textwidth]{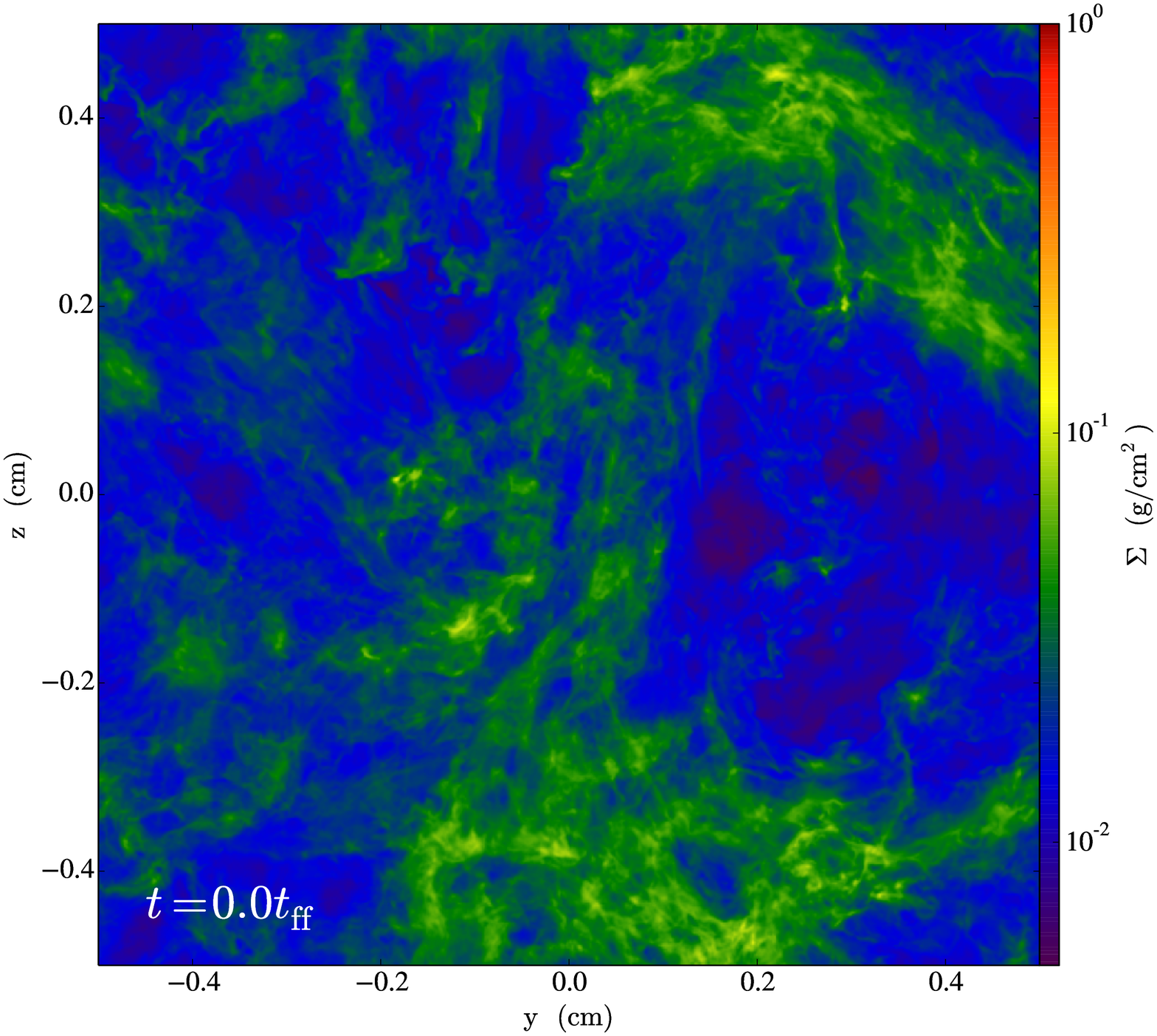}
\includegraphics[width=0.49\textwidth]{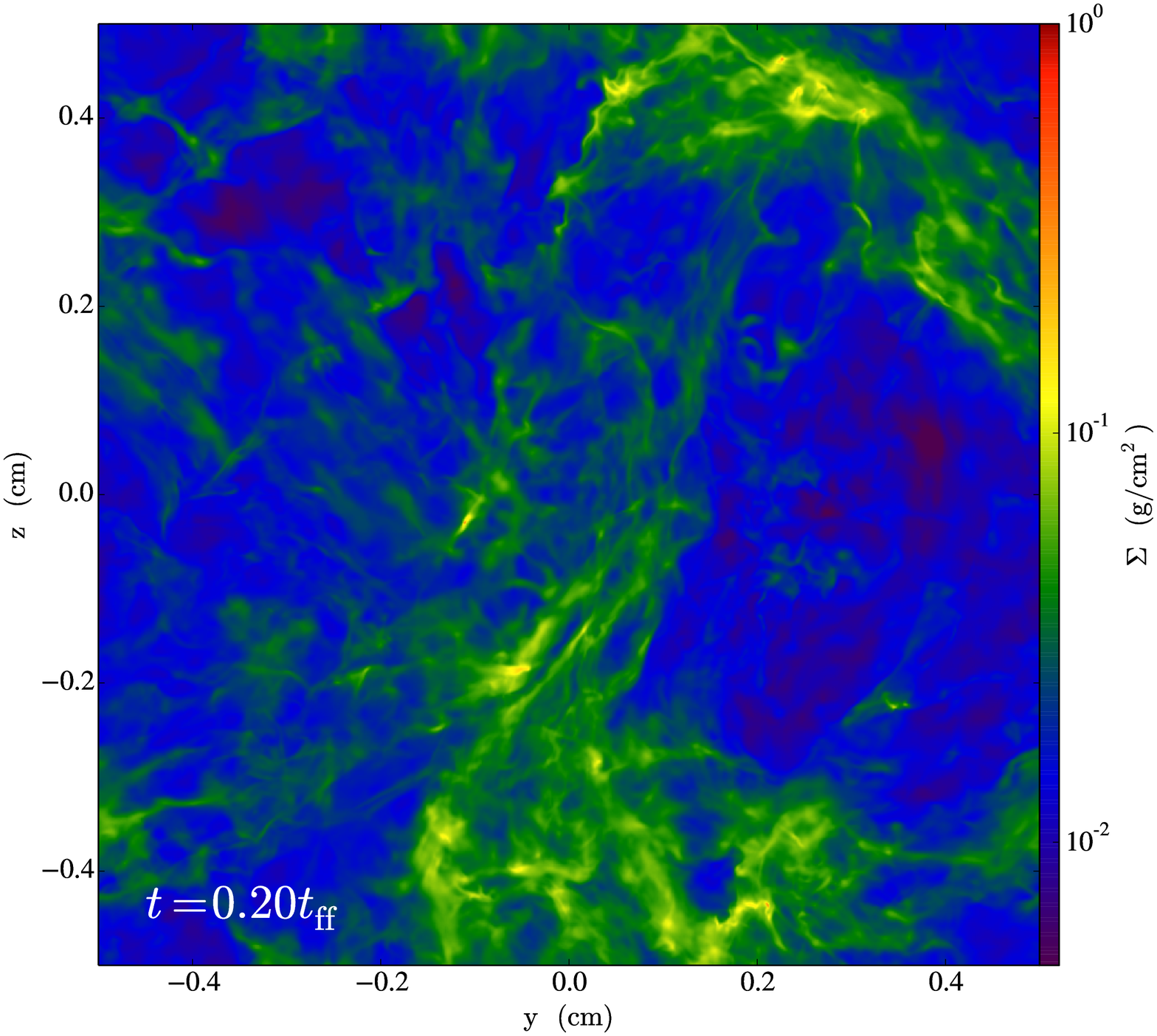}
\includegraphics[width=0.49\textwidth]{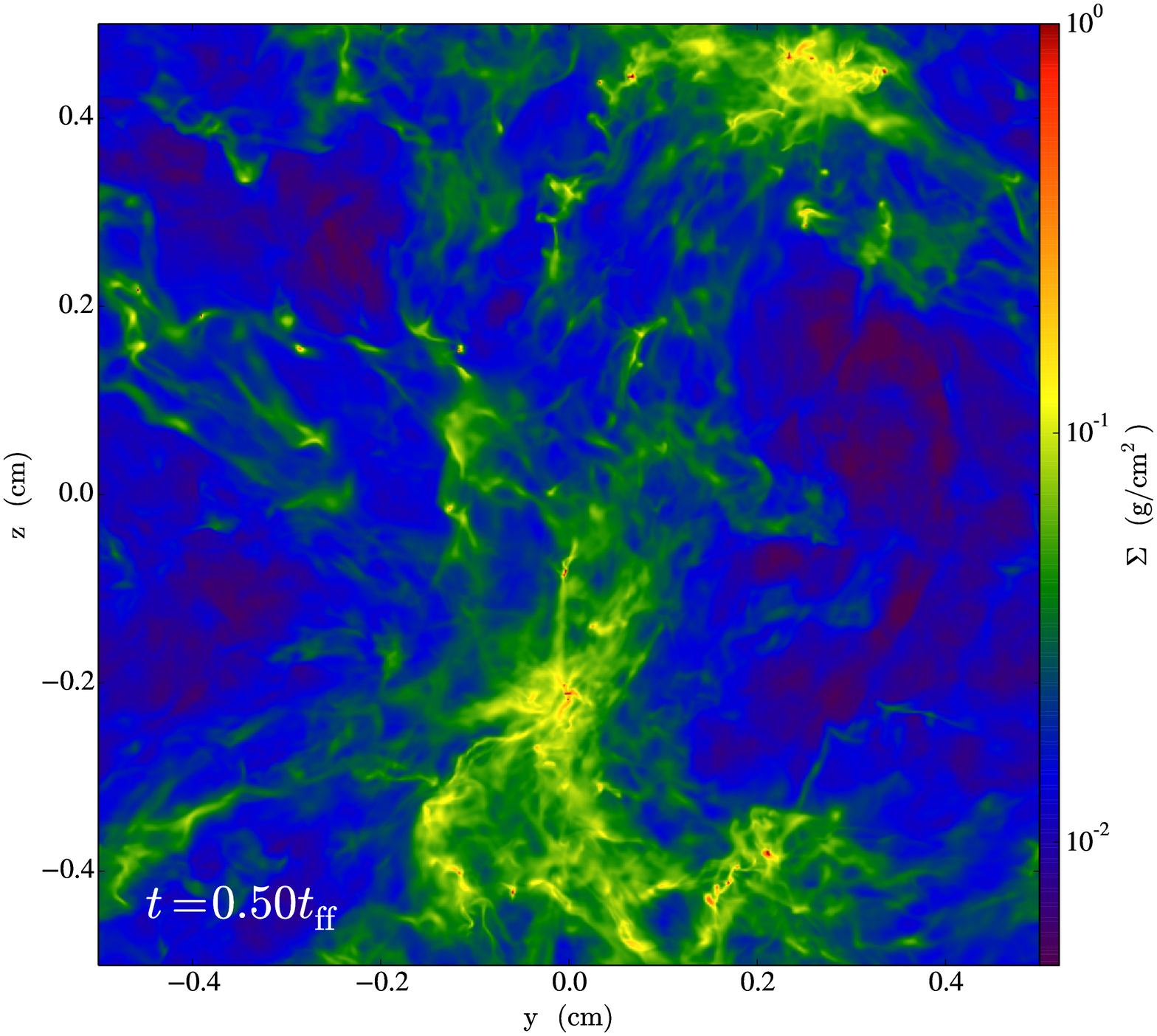}
\includegraphics[width=0.49\textwidth]{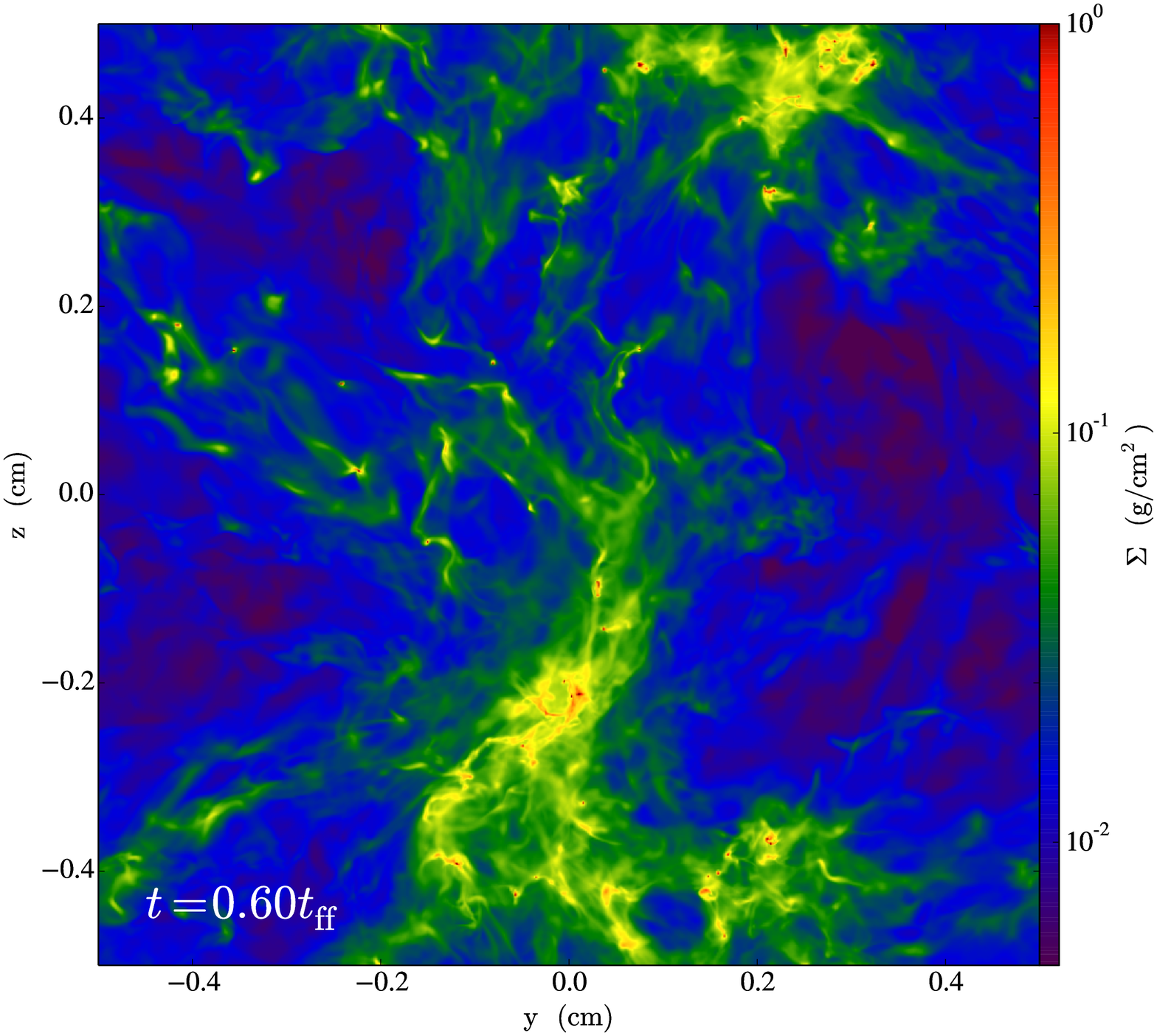}
\caption[ ]{Column density plots for the high magnetic field Enzo simulation.  Time is
shown in the bottom left.}
\label{fig:images} \end{center} \end{figure*}

In all simulations, the ideal MHD equations are solved in a periodic box, using
an isothermal equation of state ($p = c_s^2\rho$) and a variety of sonic and
Alfv\'enic Mach numbers (${\cal M}_s = \langle|\vvec|\rangle/\langle c_s\rangle$
and ${\cal M}_A=\sqrt{\rho} \langle |\vvec|\rangle/\langle |\Bvec| \rangle$,
respectively).   Here, $\rho$ is density, ${\bf
v}$ is velocity, ${\bf B}$ is magnetic field, $p$ is the gas pressure, and $c_s$
is the isothermal speed of sound.  
In all simulations, periodic cubes with initially uniform density and magnetic
fields are driven with solenoidal forcing, using the ideal MHD equations.  The
self-gravitating simulations were driven until a steady state was reached, at
which point gravity was turned on.  Driving continued during the collapse phase.  
A summary of the simulations can be found in
Table \ref{tab:models}.
Four representative snapshots of the Enzo simulations can be
seen in Figure \ref{fig:images}.

\begin{table*}
\begin{center}
  \caption{A summary of the simulations presented here.
\label{tab:models}}
\begin{tabular}{cccc}
\hline\hline
Model  & ${\cal M}_s$ & ${\cal M}_A$ &Description \\
\tableline
1  &0.5 &0.7 & subsonic \& sub-Alfv\'enic, Godunov    \\
2  &4.2 &0.7 & supersonic \& sub-Alfv\'enic, Godunov   \\
3  &8.0 &0.7 & supersonic \& sub-Alfv\'enic, Godunov   \\
4  &10 &0.7 & supersonic \& sub-Alfv\'enic, Godunov   \\
5  &20 &0.7 & supersonic \& sub-Alfv\'enic, Godunov     \\
6  &0.5 &2.0 & subsonic \& super-Alfv\'enic, Godunov   \\
7  &4.2 &2.0 & supersonic \& super-Alfv\'enic, Godunov \\
8  &8.0 &2.0 & supersonic \& super-Alfv\'enic, Godunov   \\
9  &10 &2.0 & supersonic \& super-Alfv\'enic , Godunov  \\
10 &20 &2.0 & supersonic \& super-Alfv\'enic,  Godunov \\
high & 8.5 & 6.5 & supersonic \& super-Alfv\'enic, Enzo \\
mid & 8.2 & 12. & supersonic \& super-Alfv\'enic, Enzo \\
low & 9.1 & 30 & supersonic \& super-Alfv\'enic, Enzo \\
\hline\hline
\end{tabular}

\end{center}
\end{table*}

 \def\csunits{\ensuremath{c_{\rm{s},2}}}
\def\perrho{\ensuremath{n_{\rm{H},3}}}
For the Enzo simulations we select the Mach 
number, ${\cal M}_s $, virial parameter, $\avir$, and mean thermal-to-magnetic
pressure ratio, $\betao$ as
\begin{align}
    {\cal M}_s  &=\frac{v_{\rms}}{\cs}= 9\\
    \avir &= \frac{5 v_{\rms}^2}{3 G \rho_0 L_0^2} = 1\\
    \betao&=\frac{8 \pi \cs^2 \rho_0}{B_0^2} = 0.2, 2, 20,
\end{align}
where $v_{\rms}$ is the rms velocity fluctuation,
$\rho_0$ is the mean density, $L_0$ is the size of the box, and $B_0$ is the
mean magnetic field.
\hltwo{
These can be scaled to physical clouds as
\begin{align}
    \tff &= 1.1  \perrho^{-1/2} \rm{Myr}\\
    L_0 &= 4.6 \csunits \perrho^{-1/2} \rm{pc}\\
    v_{\rms} & = 1.8 \csunits \kms \\
    M &= 5900 \csunits \perrho^{-1/2} \msun\\
    B_0 &= (13, 4.4, 1.3) \csunits \perrho^{1/2} \mu \rm{G},
\end{align}
where $\csunits=0.2\kms$ and $\perrho=n_H/(1000 \percc)$ are the sound speed and
hydrogen number density, respectively, and we have used a mean molecular weight
of 2.3 amu per particle.}

In what follows we use column density maps from the Enzo and Godunov simulations to investigate the utility
of often used statistical descriptors of turbulence and gravitational collapse, including the PDFs, Tsallis distribution
and Fourier power spectrum, for observations of column density, e.g. dust extinction maps. 
We account for fluctuations along different sight-lines relative to the orientation of the mean field in
error bars, taken as the standard deviation between a measure along different sight-lines.  
As observers often do not have information regarding the mean field we treat the line of sight relative to the mean magnetic field as 
an unknown parameter.  
Regarding the ENZO simulations, they are all super-Alfv\'enic and thus any anisotropy introduced by the 
mean field will be very weak (see \citet{Burkhart14}). The Godonov simulations do have sub-Alfv\'enic 
snapshots, however the effect of anisotropy in these is studied in other works
\citep[see][]{Esquivel11, Burkhart14}).

\section{PDFs}
\label{sec.columndensity}
The Probability Density Function of gas and dust in star forming regions provides important signatures of both MHD turbulence and gravitational collapse.
In the case of low density MHD turbulence, where self-gravitational influences are negligible, the PDF exhibits a lognormal form for both the density and column density distributions, i.e.
 \begin{align}
V(\rho) d\ln \rho = \frac{1}{\sqrt{2 \pi \sigma^2} }~ \rm{exp}\left( \frac{( \ln \rho -
    \mu)^2}{2 \sigma^2} \right ) d\ln \rho,
\label{eqn.lognormal}
\end{align}
where $\mu =
-\sigma^2/2$ is the mean of $\ln \rho$ and $\sigma$ is the standard deviation \citep{Blaisdell93,Vazquez-Semadeni94}.

Lognormal column density PDFs have been observed in various phases of the ISM including in molecular 
\citep{Vazquez-Semadeni94, Brunt10, Kainulainen11, Burkhart12, Molina12, Kainulainen13} and in the diffuse warm neutral and ionized ISM \citep{Hill08, Burkhart10}. 
At the onset of gravitational collapse the shape of the turbulence induced lognormal PDF begins to become skewed toward the high density material
which manifest as a power-law tail \citep{Klessen00, Collins12}.
Furthermore, the PDF was shown to be important for analytic models of star formation rates and initial mass functions 
\citep{Padoan02, Krumholz05, Hennebelle08c, Padoan11, Federrath12, Federrath13}.

What can be learned from the lognormal (i.e. turbulence dominated) portion of the PDF?
Several authors have suggested the turbulent sonic Mach number 
can be estimated from the calculation of the density/column density variance \citep{Padoan97b, Price11}
and the density/column density skewness and kurtosis, i.e. higher order moments \citep{Kowal07, Burkhart09}.

In particular, the relationship between  ${\cal M}_s$ and the variance of the logarithm of the 3D density distribution 
as seen in numerical models \citep{Padoan97b, Passot98} generally takes the form: 
\begin{equation}
 \sigma_{\rho/\rho_0}^2=b^2{\cal M}_s^2
\label{eq:linvar}
\end{equation}
where $\rho_0$ is the mean value of the 3D density field, $b$ is a constant which 
depends on the driving of the turbulence in question with $b=1/3$ for solenoidal forcing and $b=1$ for compressive driving, and $\sigma$ is the standard deviation of the density field
normalized by its mean value \citep{Nordlund99, Federrath08, Federrath10, Molina12}. 

When taking the logarithm of the normalized density field this relationship becomes:
\begin{equation}
 \sigma_{s}^2=\ln(1+b^2{\cal M}_s^2)
\label{eq:logvar}
\end{equation}
where $s=\ln(\rho/\rho_0)$ and $\sigma_{s}$ is the standard deviation of the logarithm of density (not to be confused with $\sigma_{\rho/\rho_0}$).

The relationship between the \textit{observable} column density standard
deviation and sonic Mach number \citep{Burkhart12} retains the same form as that of the 3D density field but with a scaling constant $A$:

\begin{equation}
\sigma_{\zeta}^2=A\ln(1+b^2{\cal M}_s^2)
\label{eq:logvar_cd}
\end{equation}

The corresponding relation for the linear variance based on Equation \ref{eq:logvar_cd} is:
\begin{equation}
 \sigma_{\Sigma/\Sigma_0}^2=(b^2{\cal M}_s^2+1)^A-1
\label{eq:linvar_cd}
\end{equation}

 where $\zeta=\ln(\Sigma/\Sigma_0)$ and $\Sigma$ is the column density distribution
with $\Sigma_0$ denoting the mean value of the column density distribution. 

In addition, the PDF of incremental fluctuations  has been shown to be useful for studies of turbulence in the density
regimes where gravity is not dominant \citep{Esquivel10, Tofflemire11}. 
The Tsallis statistic provides a fit for incremental PDFs and the fit parameters describing the width and amplitude are related
to the physics of the gas such as the sonic and Alfv\'enic Mach numbers.

In the high density regime (i.e., A$_V > 2$) regime, where gravity dominates the PDF shape \citep{Schneider13}, power law tails form with exponent
related to the star formation efficiency \citep{Federrath13}.
These power law tails have been observed in numerical simulations of both density \citep{Collins12} and column density \citep{Federrath13} as well as observations of dust extinction in numerous star forming regions \citep{Kainulainen11, Schneider13}.

In this section we focus on the column density PDFs, Tsallis function and PDF moments in MHD turbulence simulations with and without gravity 
for a range of magnetic field strengths. In the case of the Enzo simulations, we pay particular attention to the time evolution in order
to see how the column density PDF evolves as the cloud collapses.

\subsection{Column Density PDF}

\begin{figure*} 
\begin{center}
\includegraphics[scale=.5]{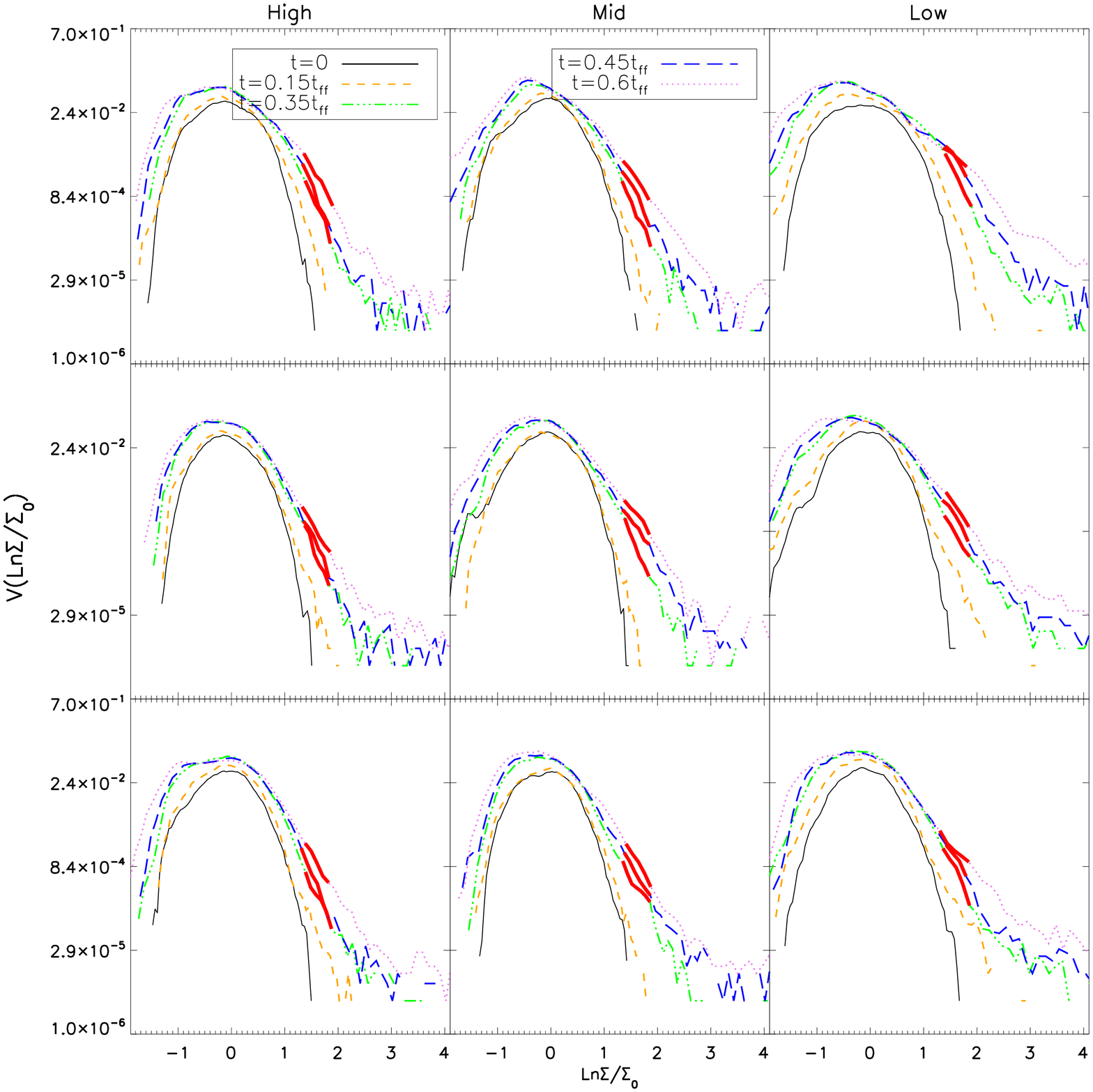}

\caption[ ]{PDFs of $\ln \Sigma/\Sigma_0$ (i.e.  $\zeta$) at $512^3$ resolution. The top row
displays line of sight (LOS) in the $x$-direction (i.e. the direction of the mean
magnetic field), the middle row displays the LOS in the $y$-direction, while the
bottom row displays the LOS in the $z$-direction. Columns
show the high, mid and low $B_0$, respectively.  Different time steps (in units
of $t_{\rm{ff}}$) are shown with the color and linestyle given in the legend.  
We
overplot the density regime  
that we fit the power law tails as thick solid lines for $t > 0.35t_{ff}$.}
\label{fig:pdfs} 
\end{center} 
\end{figure*}

Figure \ref{fig:pdfs} shows the column density PDFs of the Enzo simulations for different line-of-sight (LOS) orientations (rows across) and different time steps (in units of $t_{ff}$)  
indicated in different colors/linestyles, with the $t=0$ case indicating pure supersonic MHD turbulence
plotted as a solid black line. The top row
displays LOS in the $x$-direction (i.e. the direction of the mean magnetic field),
the middle row displays the LOS in the $y$-direction, while the bottom row
displays the LOS in the $z$-direction.

We fit lognormals to the peaks of the distribution 
and separately fit a  power law  for $t > 0.35t_{ff}$ in the column density range of $\zeta=1.3-1.9$.  This fit range avoids the very high
density portion of the PDF, which is plagued by low number statistics, as well as the peak of the distribution. 
This range is denoted by the thicker red line in plot.
We divide the PDFs into three different time regimes and point out the following visual features:
\begin{enumerate}
\item At $t\leq 0.15t_{ff}$, PDFs (black solid and yellow dashed lines) display general lognormal behavior.  The simulation at $t=0.15t_{ff}$ shows a wider log normal than 
the $t$=0 simulation for roughly the same sonic Mach number. For $t<0.15t_{ff}$ in the
high density range we do not observe a clear power law tail or the tail is very steep such that it is visually
indistinguishable from the lognormal 
\item At $0.35t_{ff} \leq t \leq 0.45t_{ff}$, the PDFs retain the general log normal shape for log column density values  $ \zeta <1$ regardless of LOS orientation or magnetic field strength,
however for column densities $\zeta > 1$ power law tails are clearly seen for time steps at $t\geq 0.35t_{ff}$
\item At $t >0.45t_{ff}$, the PDFs (purple dotted lines) show clear power law tails at the high column density end while the PDF distribution of low column density material has become broadened
as compared with earlier times.  This broadening of the low column density
material has also been observed in density (see \citet{Collins12}).
\end{enumerate}

Examining the PDFs of Figure \ref{fig:pdfs} also shows that there are differences in the behavior for different LOS orientations 
and that the magnetic field strength plays a role in the PDF shape and onset of the power law tail.  
In particular,  turbulence with a lower magnetic field (far right column denoted with \textit{low} in the title of the plot) 
exhibits wider PDFs at low densities and more pronounced and shallower power law tails at the higher densities.  
Both the wider PDFs and shallower power law tails are more exaggerated when the LOS is parallel (top row) to the 
magnetic field rather then perpendicular (middle and bottom rows).  We investigate these effect quantitatively in the following subsections
by exploring the PDF moments and Tsallis statistics as well as fitting the power law tail exponential, $\alpha$.

\subsection{Power Law Tail}
\label{sec.powerlaw}
The PDF of the collapsing matter forms a power law, i.e. $V(\rho) \approx \rho^{\alpha}$,
for densities above a critical density \citep{Klessen00,Slyz05,Vazquez08}. 
High resolution AMR simulations by 
\citet{Kritsuk11} measured a range of slopes:  -1.67 at intermediate
densities and -1.5 at high densities. They posit that the flattening of the
slope is due to the onset of rotational support, which is backed up by the
analysis of the support function of that data by \citet{Schmidt13}.
Recently, \citet{Federrath13} investigated the power law tail indices of 
self-gravitating MHD simulations and found that the high density tails are consistent with equivalent radial density profiles, $\rho \propto r^{-\kappa}$
with $\kappa=1.5-2.5$.  
Observational constraints of the power law tail of the column density
PDF are comparable with ranges reported by simulations \citep{Arzoumanian11,Schneider13}.


\begin{figure*} \begin{center}
\includegraphics[scale=0.5]{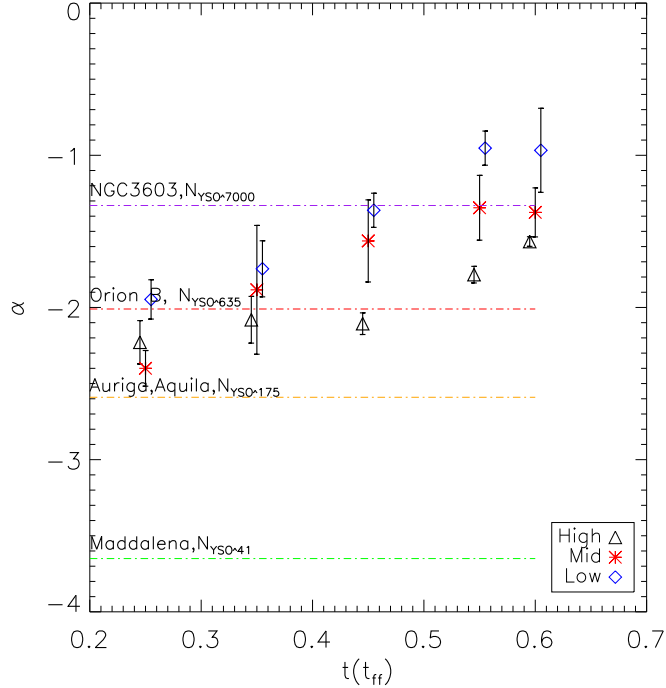}
\caption[ ]{The power law tail index ($\alpha$) of the  $512^3$ resolution Enzo
simulations, averaged over all LOS, versus time evolution. Error
bars denote the standard deviation between different LOS.  Lines indicate
observed powerlaws from \citet{Schneider13b} and \citet{Schneider14}
 }
\label{fig:alpha2} \end{center} \end{figure*}

We plot the power law tail exponent $\alpha$ versus time evolution for  $t > 0.2t_{ff}$ averaged over the three cardinal LOS directions with error bars indicating the standard deviation between 
different LOS in Figure \ref{fig:alpha2}. 
The LOS relative to the mean magnetic field is generally an unknown parameter for observational studies.  
We fit the slope to the range of column densities show in Figure \ref{fig:pdfs} (i.e. $\ln \Sigma/\Sigma_0=1.3-1.9$).
These values were chosen to avoid both the low number statistics
of the high density portion of the PDF and the peak of the PDF.  We will address a robust fit for the power law tails
based on maximum-likelihood fitting in a future work.

We include the published values for the power law tail slopes and approximate
number of young stellar objects (YSOs) for  five
molecular clouds from Herschel data from \citet{Schneider13b, Schneider14}: NGC3603, which is an active low-mass star forming region the Auriga cloud,
 as well as the Orion B, Maddalena, and Aquila clouds.
We discuss these clouds and their relation to the simulations in the discussion section.  We omit the PDF
estimates from the Carina cloud  as the dynamics of Carina may include significant massive stellar feedback
which is not treated in our simulations.
This comparison is complementary to a
study by \citet{Kainulainen14}
 who showed that a proxy for the 3D density PDF could be related to the number of YSOs in several different clouds.

The power law tail index is steepest for earlier time steps and generally becomes increasingly shallow as 
the cloud proceeds to collapse. This is expected from past studies of the column density power law tail index 
and is also observed in the density PDF \citep{Kritsuk11, Collins12,Federrath12, Federrath13}.
In addition to the shallowing of the power law index with time, Figure \ref{fig:alpha2} shows that
there is a strong dependency on the magnetic field.   Simulations with higher field strengths show steeper slopes for all times steps due to
the suppression of density enhancements by the magnetic field.
The low magnetic field run shows shallower values of $\alpha$ across the time evolution parameter space.

The low magnetization simulation  has values of $\alpha$
ranging from -2.5 to -0.7, which falls out of the bounds set by the observations at later times.
This simulation has an Alfv\'enic Mach number $\approx 40$, which may be two high for realistic clouds. 
This may suggest super-Alfv\'enic clouds have Alfv\'enic Mach numbers in the range of 6-12, which is representative of our
high and mid magnetic field simulations and has been argued for in a number of past works
\citep{Padoan99, Lunttila08, Burkhart09, Crutcher09, Collins12}.

\subsection{Column Density PDF Variance}
\label{sec.variance}

\begin{figure} \begin{center}
\includegraphics[scale=.45]{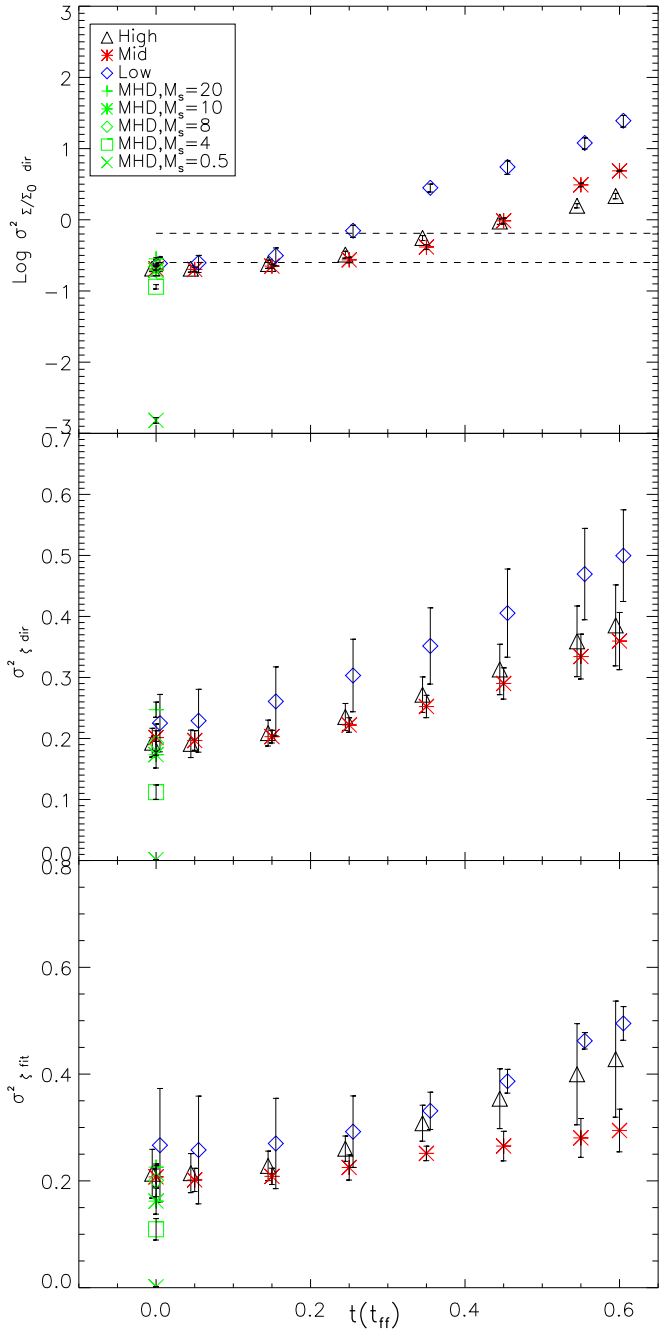}
\caption[ ]{Different methods of calculation of the PDF variance versus time. The
Godunov  MHD turbulence simulations are plotted with green symbols at $t$=0.
Top row: directly calculated variance of the linear column density distribution. Straight
dashed lines across represent the range of values found in the
\citet{Kainulainen09} survey of IRDC PDFs.
Middle row:  directly calculated variance of the natural logarithm of the column density distribution.
Bottom row: Gaussian fitted variance of the column density distribution.
Error bars are calculated as the standard deviation between three different sight lines. }
\label{fig:var} \end{center} \end{figure}

We investigate the PDF variance as a function of time evolution in Figure \ref{fig:var} for the Enzo and Godunov simulations.   
We investigate several different methods of calculation of the variance in  Figure \ref{fig:var}, including
directly calculating the variance of the column density distribution ($\sigma^2_{\Sigma/\Sigma_0 dir},$ top panel) and directly calculating the variance of the natural logarithm of the column density
distribution ($\sigma^2_{\zeta dir}$, center panel).

For a distribution $\Sigma={\Sigma_i, i=1...N}$  the mean value is defined as:
$\Sigma_0=\frac{1}{N}\sum_{i=1}^N {\left( \Sigma_{i}\right)}$.  We define the direct calculation of the variance as:
\begin{equation}
\sigma^2_{\Sigma/\Sigma_0 dir}= \frac{1}{N-1} \sum_{i=1}^N {\left(
\frac{\Sigma_{i}}{\Sigma_0} - {\left( \frac{\Sigma_{i}}{\Sigma_0}\right )_0}\right)}^2
\end{equation}
The calculation of the variance directly from the data does not assume that the PDF follows a particular model (i.e. that it is a lognormal).
We also calculate directly the variance of natural logarithm of the column density distribution as:
\begin{equation}
\sigma^2_{\zeta dir}= \frac{1}{N-1} \sum_{i=1}^N {\left( \zeta_{i} - {\zeta_0}\right)}^2
\end{equation}

Figure \ref{fig:var}, bottom panel, shows the variance calculated by fitting a Gaussian to the peak of the distribution and measuring the variance from the fit
($\sigma^2_{\zeta fit}$).

The column density variance of the MHD simulations at $t=0$ depends primarily on
the sonic Mach number \citep{Burkhart12} regardless of the
calculation method.
The distribution of gas with larger sonic Mach number shows increasing variance for both direct calculation of the variance
and the variance calculated from a Gaussian fit.  When gravity begins to alter
the PDF from lognormal, i.e. at  $t > 0$, 
the variance ($\sigma^2_{\zeta dir}$ and $\sigma^2_{\Sigma/\Sigma_0 dir}$) of the 
Enzo simulations (which have sonic Mach numbers of $\approx 10$) dramatically increases past the expectations of MHD turbulence even with Mach numbers as high as 20.  
This effect manifests as several orders of magnitude larger difference in the directly calculated variance (top panel) of the column density distribution.  The natural logarithm of the column density distribution (middle panel) also shows a linear increase in variance as the cloud evolves
with gravity.  The variance as obtained from a Gaussian fit remains generally flat within the 
error bars across the time evolution parameter space up until about $t=0.4 t_{\rm{ff}}$ for the low and mid magnetic field runs and for the
high field run the variance does not significantly change in the time parameter space investigated.
This is because the Gaussian fit is applied to the peak of the distribution and is sensitive primarily to the lower density material dominated by  turbulence and not the power law tail high
density portion of the PDF.  However, there is a slight trend toward a wider
lognormal at the low density end of the PDF (which can be visually seen in
Figure \ref{fig:pdfs}).  This suggests that fitting a Gaussian to the peak of the distribution can be used to 
dissect the turbulence dominated portions of the gas while the fraction of gas collapsing to form proto-stars may be 
probed with the formation of the power law tail index (see Figure \ref{fig:alpha2}).

Additionally the variance of collapsing column density material shows a dependency on the magnetic field. 
For all three methods of variance calculation,
the simulation with lower field strength shows higher values of variance as compared to the high/mid cases which are generally
not distinguishable within the error bars.  This effect is more pronounced and increasingly
significant across different sight lines  as the time
evolution increases. As the magnetic field increases so does the suppression of
high density clumps, which causes the column density variance to be smaller compared 
with simulations with lower field strengths. 
The dependency of the column density variance on cloud magnetization in gravoturbulence simulations is somewhat surprising as 
the variance has been shown to be weakly dependent magnetic field strength in other works that focused only on supersonic
MHD simulations without gravity \citep{Burkhart10,Molina12}. These results suggest the magnetic field plays a role in 
the global support of the cloud against gravity.

Comparison with published dust extinction column density maps from
\citet{Kainulainen13} (Table 1 of that work) shows that clouds have values of directly calculated variance
ranging from $\sigma^2_{\Sigma/\Sigma_0}=0.25-0.64$.  Taking the logarithm of
these numbers for ease of comparison with the top panel of  Figure
\ref{fig:var}, these values are log$\sigma^2_{\Sigma/\Sigma_0}=-0.6$ to $-0.19$, which suggests
the cloud variance can not be attributed to MHD turbulence alone, and that
gravity must be active up to $t=0.5t_{ff}$ as compared with our  high or mid
magnetic field simulations.  We over plot the \citet{Kainulainen13} cloud
variance ranges with straight dashed lines in the top panel of Figure
\ref{fig:var}. 

\begin{figure*} \begin{center}
\includegraphics[scale=.8]{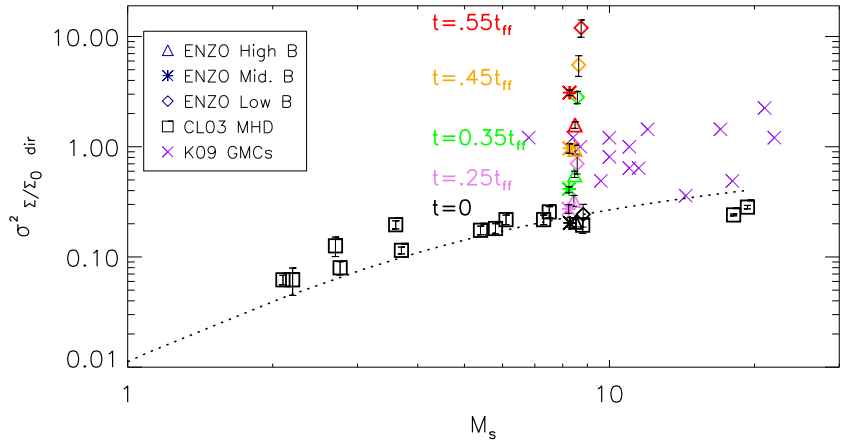}
\caption[ ]{$\sigma^2_{\Sigma/\Sigma_0 dir}$ vs ${\cal M}_s$
for the Godunov simulations and Enzo simulations.
Pink, green, orange and red colors represent increaseing time in the Enzo simulations. The black dotted lines represent 
Equation \ref{eq:linvar_cd} with $A=0.11$ and  $b=1/3$. Purple X symbols
represent GMCs from \citet{Kainulainen09} as taken from Table 3 of \citet{Kainulainen13}.}
\label{fig:varms} \end{center} \end{figure*}

We compare the self-gravitating snapshots from the Enzo simulations 
to the column density variance - sonic Mach number relation given in Equation (\ref{eq:linvar_cd})
in Figure \ref{fig:varms}. Our general conclusions indicate that two regimes exist which are defined by the importance of self-gravity:

\begin{enumerate}
 \item The column density variance - sonic Mach number relation without gravity:
the Godunov (black squares) and  $t=0$ Enzo simulations (black symbols denoted in the legend) follow closely the prediction
of equation \ref{eq:linvar_cd} given by \citet{Burkhart12}.

\item The column density variance - sonic Mach number relation with gravity:
once gravity  becomes important  the PDF variance no longer tracks the behavior
of the sonic Mach number. The more evolved the cloud's collapse, the higher the variance even for the 
same sonic Mach number. 

\end{enumerate}

Figure \ref{fig:varms} highlights the importance of the magnetic
state of the gas in the evolution of the PDF of GMCs. 
The variance of the Enzo simulations depends strongly on the global strength of the magnetic field.  
The lower the value of the magnetic field, the higher the measured variance.  This is because simulations
with stronger field suppress the formation of dense cores as a function of time.

Figure \ref{fig:varms} also compares the Enzo and Godunov column density variance - sonic Mach number relation with that of 
the GMCs from \citet{Kainulainen09} as taken from Table 3 of \citet{Kainulainen13}.
The GMC variance values are all larger than the Godunov simulations and follow
the upward evolution of the Enzo simulations along the variance axis.   A direct comparison of the range of variance values
with clouds that have $M_s \approx 10$ suggests evolution time scales of $t=0.25-0.45t_{ff}$ over a range of magnetic field values.


\subsection{Column Density PDF Higher Order Moments}

\label{sec.moments}
\begin{figure} \begin{center}
\includegraphics[scale=0.4]{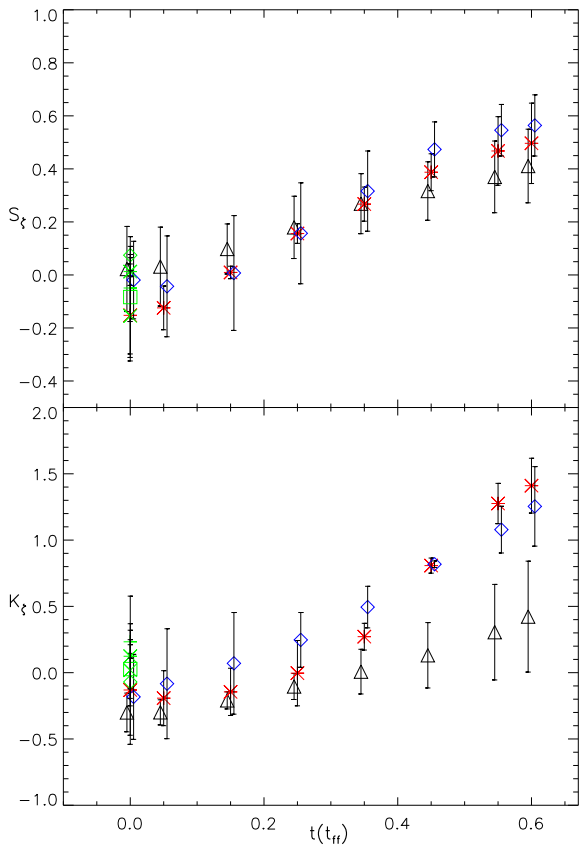}
\caption[ ]{Higher order moments vs. time evolution of the natural logarithm of the column density distribution ($\zeta$).  
Skewness ($S$) is shown in the top panel and kurtosis ($K$) is shown in the bottom panel.
We use the same color scheme as Figure \ref{fig:var}.
Later times show higher skewness and kurtosis but unlike the variance, little dependency magnetic field variation is found. }
\label{fig:high} \end{center} \end{figure}

In addition to the variance, the higher order moments of the PDF have also been shown in a number of works to be sensitive diagnostics of the sonic Mach number of a turbulent density field \citep{Kowal07, Burkhart09, Burkhart10}.  However these earlier works investigated the higher order moments of the linear column density distribution (i.e. $\Sigma/\Sigma_0$).  As shown in the previous subsections, the power law tail of the column density distribution is a sensitive diagnostic of the evolutionary state of a collapsing cloud.  A additional sensitive diagnostic that could be complimentary to the power law tails seen in  the natural logarithm of the column density distribution ($\zeta$) are the higher order moments  precisely because they characterize deviations from Gaussianity.

Skewness and kurtosis are defined by the third and fourth-order statistical moment. Skewness is defined as:
\begin{equation}
S_{\zeta} = \frac{1}{N} \sum_{i=1}^N{ \left( \frac{\zeta_{i} - \overline{\zeta}}{\sigma_{\zeta}} \right)^3 }
\label{eq:skew}
\end{equation}
If a distribution is Gaussian, the skewness is zero.   Kurtosis is a measure of whether a quantity has a distribution that is peaked or flattened compared to a normal Gaussian distribution.  If a data set has positive kurtosis then it will have a distinct sharp peak near the mean and have elongated tails.  If a data set has negative kurtosis then it will be flat at the mean. Kurtosis is defined as:
\begin{equation}
K_{\zeta}=\frac{1}{N}\sum_{i=1}^N \left(\frac{\zeta_{i}-\overline{\zeta}}{\sigma_{\zeta}}\right)^{4}-3
\label{eq:kurt}
\end{equation}

We plot the higher order moments of the natural logarithm of the column density distribution in Figure \ref{fig:high}.
The expectations for MHD turbulence with no gravity (i.e. our $t=0$ cases) are
values of skewness and kurtosis bounded around zero as the lognormal distribution should have relatively small skewness and kurtosis for given 
sonic Mach number\footnote{This is not the case for the linear distribution of density. 
For example, \citet{Burkhart09} showed that the skewness and kurtosis are sensitive to sonic Mach number 
for linear density and column density as shocks increase tails toward the high density portion of the PDF}.  An increase in sonic Mach number affects the width of the lognormal PDF
and generally not the tails or peaks.  Indeed, at $t<0.2t_{ff}$  the values of skewness are bounded between 
0.4 and -0.4 and the values for kurtosis are bounded between -0.6 and 0.8.  
For $t>0.2t_{ff}$ the values of skewness and kurtosis are generally positive and  increase with time evolution 
due to the formation of the power law tail that creates more peaked distributions and skews the PDF tail toward the high density end.
In general, we do not see a systematic sensitivity to the magnetization of the simulations, however the low
and mid simulations generally show higher values of skewness and kurtosis than the simulation with high value of magnetic field.

\begin{figure*} \begin{center}
\includegraphics[scale=.6]{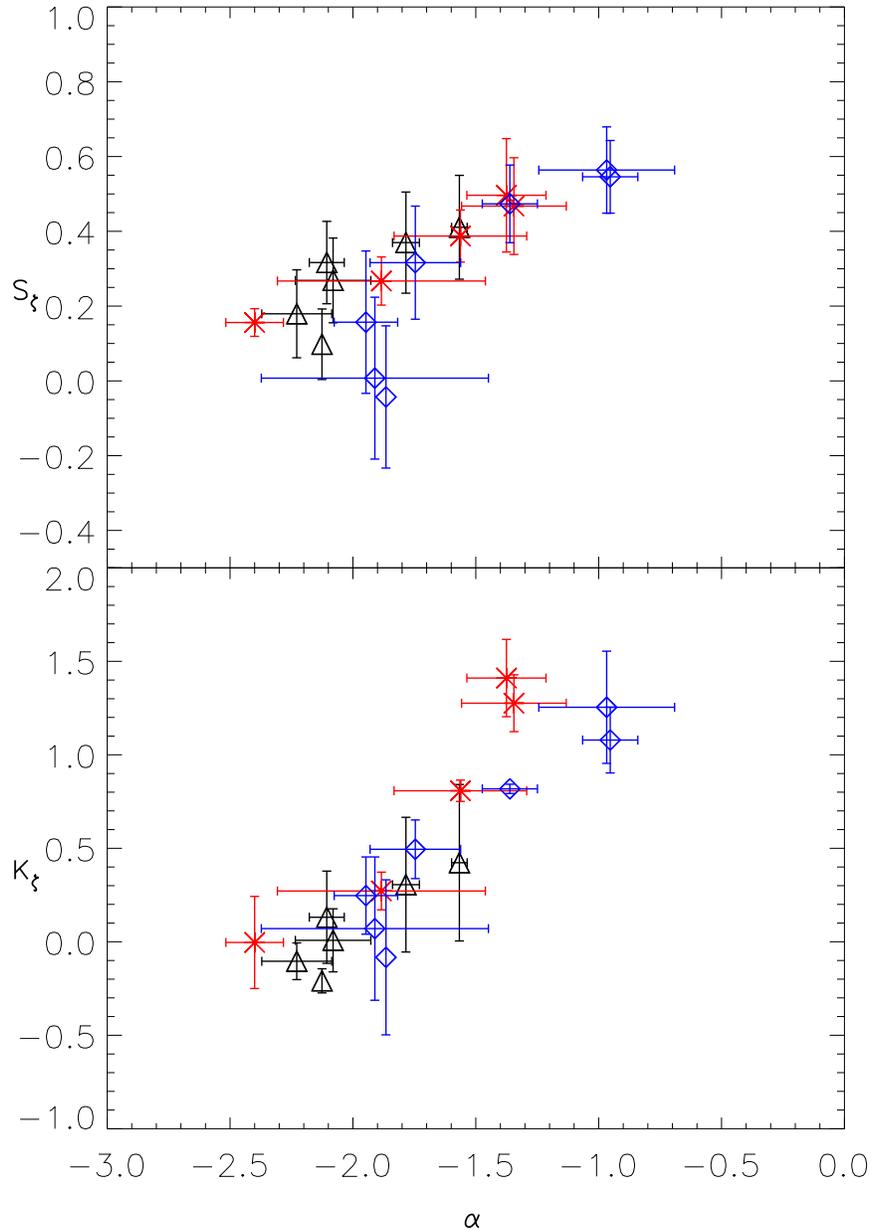}
\caption[ ]{Power law index $\alpha$ vs. skewness (top panel) and kurtosis (bottom panel).
We use the same color scheme as Figure \ref{fig:var} for the Enzo simulations.}
\label{fig:tailvskew} \end{center} \end{figure*}

Figure \ref{fig:high} suggests that calculation of the  skewness and kurtosis of the natural logarithm of the column density distribution (i.e. $\zeta$)
can be complimentary to fitting the power law tails to the high density material. 
We test this idea by plotting the power law tail index $\alpha$ vs. the skewness and kurtosis of $\zeta$ in Figure \ref{fig:tailvskew}.
Both the higher order moment and the power law tail index increase monotonically with time evolution  and hence can be seen to increase together.
Although the power law tails have dependency on the magnetization of the gas (see Figure \ref{fig:alpha2}), the higher order moments of $\zeta$ are somewhat insensitive to
field strength.  This is clear in that the lowest magnetic field run (blue data) proceeds out to much higher values of $\alpha$ for the same spread in values of
skewness/kurtosis.  This implies that the degeneracy can be broken and that  observers who measure higher order moments and power law tails out to larger values  might constrain both the energetic importance of the magnetic field as well as the time evolution of clouds.

The many avenues of exploring PDFs, including
fitting a Gaussian to the peak and a power law to the high density tail region, as well as direct calculation
of the PDF moments of the entire distribution provide researchers with a tool
kit for separating out the contributions
of MHD turbulence and gravity in the structure formation and evolution of clouds.  In the next sub-section we also suggest
spatial incremental PDFs to be of use toward this purpose.

\subsection{Column Density Incremental PDFs: the Tsallis Statistic}
\label{sec.tsallis}

In studying PDFs we study the column density field as it is given from observations. 
The density may be influenced by multiple processes that act differently on different scales.
 Thus it is advantageous to use a measure which can differentiate the properties
 of turbulence at different scales, e.g.,
 at the scales of energy injection, inertial range interval, energy dissipation scale and scales where gravitational interactions become important.
 To do this, it is advantageous to use
incremental measures that are describe the increments of densities over the scale $r$,
namely, $\rho(x+r)-\rho(x)$. Such incremental PDFs may be studied by different means and
fitting those to the Tsallis distribution is a way to do the job.

 The Tsallis distribution is often used in the context of non-extensive statistical dynamics. 
It was originally derived (see \citet{Tsallis88}) from an entropy generalization to extend the traditional
 Boltzmann-Gibbs statistics to multi-fractal systems (such as the ISM). The Tsallis distribution has since been
applied to problems over a range of applications as the distribution can describe PDFs with tails that are not
exponentially bounded.

The Tsallis function of an arbitrary incremental PDF $(\Delta f(r))$ has the form:

\begin{equation}
R_{q}= a \left[1+(q-1) \frac{\Delta f(r)^2}{w^2} \right]^{-1/(q-1)}
\label{eq:tsal}
\end{equation}

where $\Delta f(r)$ denotes the normalized PDF of incremental differences (with spatial separation $r$), i.e.
\begin{equation}
  \Delta f(r)=(f(\mathbf{x,r})-\left\langle f(\mathbf{x,r})\right\rangle_{\bf{x}})/\sigma_{f},
\end{equation}
where $\sigma_{f}$ the dispersion of the increments and $\langle...\rangle_{\bf{x}}$ denotes
spatial averaging over a shell of size $r=|\bf{r}|$.
The normalization used is such that the PDF has mean value at zero and a standard deviation of unity.
We denote the column density increments as: 
$f(\bf{x, r})=\Sigma(\bf{x + r}) - \Sigma(\bf{x})$,
A normalized histogram of
the incremental maps for a given lag results in our incremental PDF which is then fit
with the Tsallis function.

The other parameters in Eq. (\ref{eq:tsal})
are as follows: $q$ is the so called ``entropic index''
 or ``non-extensivity parameter'' which is related to the size of the tail of
 the distribution;
$w$ is a measure of the with of the PDF related to the width of the
distribution;
and $a$, the amplitude. 
By varying the parameter $q$ in the Tsallis distribution it is possible to obtain distributions 
that range from Gaussian at $q\rightarrow 1$ to ``peaky'' distributions with large tails. 
The parameter $q$ is closely related to the kurtosis (fourth order one-point moment) 
of the PDF, and similarly the parameter $w$ is related to the variance of the PDF.

The Tsallis distribution
 reduces to the classical Boltzmann - Gibbs (Gaussian) distribution in the limit of
$q\rightarrow 1$. However, for the present purpose we use it as an empiric function that
fits well the properties of MHD turbulence as was shown in \citet{Esquivel10}
(see also \citet{Tofflemire11}). We also note that the Tsallis function was
successfully used to characterize the magnetic field of the solar wind in a series of papers
by \citet{Burlaga04, Burlaga04b, Burlaga05, Burlaga05b, Burlaga06, Burlaga07,
Burlaga09}.
\citet{Esquivel10} and \citet{Tofflemire11} used the Tsallis statistics to
describe the spatial variation in PDFs of turbulent density, column density,  velocity, and magnetization of
 MHD simulations without gravity.

\citet{Esquivel10} and \citet{Tofflemire11} used the Tsallis statistics to
describe the spatial variation in PDFs of turbulent density, column density, velocity, and magnetization of
 MHD simulations without gravity. Both efforts
found that Tsallis provided adequate fits to their incremental PDFs and gave insight into
statistics of MHD turbulence. Our present study is the first attempt to apply the Tsallis
statistics to the density field obtained with the simulations that include self-gravity.

For non self-gravitating MHD simulations it was shown in \citet{Esquivel10} and \citet{Tofflemire11}
that the Tsallis fit parameters $a$, $q$ and $w$ of the column density distribution
 were sensitive to both the sonic and Alfv\'en Mach numbers. Higher sonic Mach numbers and higher magnetic field
strengths produced incremental PDFs with higher values of width ($w$), amplitude ($a$) and kurtotic index ($q$).
In this subsection we investigate the first use of the Tsallis function to simulations of self-gravitating MHD turbulence.

\begin{figure*}[scale=0.8] \begin{center}
\includegraphics[]{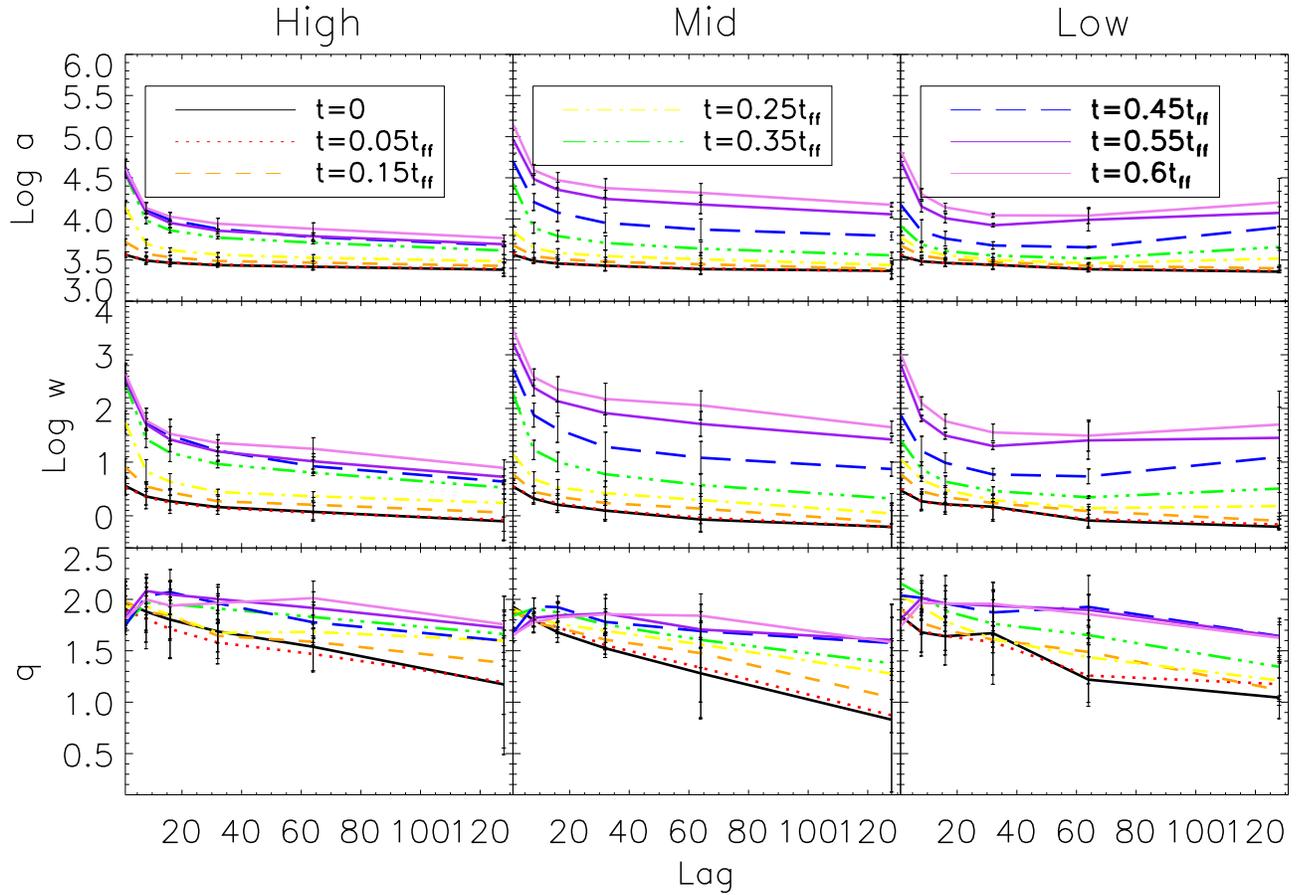}
\caption[ ]{Tsallis fit parameters vs. incremental lag.  The top row shows the
logarithm of the amplitude parameter (Log $a$), the middle row shows the 
logarithm of the width parameter (Log $w$) and the bottom row shows the kurtotic parameter ($q$).  
Columns show high, mid and low magnetic field strength simulations, respectively.
The color scheme follows that of Figure \ref{fig:pdfs} except that we include two additional models: purple solid line with $t=0.55t_{ff}$ and yellow dashed and dotted line with $t=0.25t_{ff}$.}
\label{fig:tsal} \end{center} \end{figure*}

Figure \ref{fig:tsal} shows the three Tsallis parameters, $a$, $w$ and $q$ vs. spatial lag (in pixels) for the Enzo simulations. 
 Error bars are plotted by taking the standard deviation of values of the fit parameters with different LOS
relative to the mean magnetic field. High, mid, and low values of magnetic field are presented for comparison
in the columns going from left to right, respectively. In simulations with and without self-gravity, signs of the
dissipation scales can be seen at small lag increments.

The solid black line represents the Enzo simulation with pure supersonic MHD turbulence. Comparison of these values with Figure 8 from \citet{Tofflemire11}
show very good agreement with other supersonic MHD simulations.
Once self-gravity begins to create regions of over-density, all three Tsallis parameters 
dramatically increase well past the expectations for supersonic turbulence.

The values of the amplitude and width of the Tsallis column density PDFs show
dependencies on the magnetization of the gas. For the low and mid magnetic field simulations, the incremental PDFs are wider and have
higher amplitudes compared with the high magnetization case. This is in contrast to the findings of \citet{Tofflemire11}, which found
that a higher field increases the Tsallis parameters in MHD turbulence without gravity.
However, when the material collapses the magnetic field acts to suppress overdense regions from forming which in turn
creates incremental PDFs with lower values of width and amplitude.
Within the error bars given by the LOS orientation the effect of the magnetic field
is not distinguishable between the mid and low magnetic field simulations but is nearly an order of magnitude different
comparing these two regimes with the high magnetic field simulation at later times (i.e. $t>0.45t_{ff}$).

The kurtotic index $q$ shows little clear variation with changing magnetic field but does show a dependency on the time evolution
of the column density. As gravity acts to create the dense clumps seen in Figure 1, the kurtotic index $q$ is seen to increase past
the turbulence only snapshot (black line).  The increase in $q$ is monotonic with increasing time step.  
The fact that $q$ does not depend strongly on magnetic field (or sonic Mach number, see \citep{Tofflemire11}) but rather only on the collapse
evolution suggests that this parameter might be extremely useful in breaking the degeneracy between gravity, compressibility and magnetization in 
the PDF statistics of column density.  We discuss this further in section \ref{sec.discussion}.

\section{Column Density Power Spectra}
\label{sec.spectra}
Complementary to PDFs, an essential tool for turbulence studies is the spatial density and velocity power spectrum.
The turbulence energy transfer
process can be studied by examining the Fourier power spectrum, and the sources and sinks of energy, including the injection and dissipation scales, can be identified. 
The power spectra of density and velocity (and their variants such as
the structure function and delta variance) have been suggested by several authors to provide information on the spatial and kinematic scaling of turbulence, 
sonic Mach number and injection/dissipation scales  \citep{Kowal07, Burkhart10, Ossenkopf01, Collins12, Federrath13}.  
In this section we explore the use of the spatial power spectrum of column density maps
of self-gravitating MHD turbulence for determining the dynamical evolution of clouds undergoing collapse and investigate if the power spectrum might be a 
complimentary tool to the PDF.  We also investigate the origin of the changes in the column density power spectrum as a function of time.

The Fourier transform of the two point autocorrelation function (i.e. the spatial power spectrum)
provides information on the properties of the turbulence cascade, including the injection scales and dissipation scales. 
The power spectrum is defined as:
\begin{equation}
P(\kvec)=\sum_{\kvec=const.}\tilde{F}(\kvec)\cdot\tilde{F}^{*}(\kvec)
\end{equation}
where $k$ is the wavenumber and $\tilde{F}(\kvec)$ is the Fourier transform of the field under study, which 
for our purposes is the synthetic column density maps. 

The one-dimensional energy spectrum $E(k)$ is related to the measured power spectrum by $E(k)dk \propto P(k)dk^D$, where D is the dimensionality.
For incompressible turbulence, the Kolmogorov power spectrum scaling \citep{kolmogorov41a}  in three dimensions (3D) 
is $P(k)_{3D}\propto k^{-11/3}$ and the energy spectrum scales as $E(k)\propto  k^{-5/3}$.
For the same $E(k)$, $P(k)_{2D}\propto k^{-8/3}$, and in 1D $P(k)_{1D}\propto k^{-5/3}$.

Although the Kolmogorov slope is for incompressible unmagnetized fluids, the analysis
of \citet{Goldreich95} showed that the energy spectrum scaling of the incompressible cascade perpendicular to the mean magnetic field
also retains the $-5/3$ slope.  The \citet{Goldreich95} analysis was extended in \citet{Lazarian99} and \citet{Cho00} 
to include the concept of the cascade relative to the 
local magnetic field to obtain the correct scaling relations. 
Later studies confirmed the $-5/3$ slope with higher resolution simulations \citep{Beresnyak09, Beresnyak12}.
The actual spectrum of MHD turbulence is anisotropic,
and scale-dependent anisotropy in the system of reference is
connected with the local magnetic field (see the discussion of the
concept of the local magnetic field, e.g. see \citet{Cho03}).
The statistics of fluctuations given by the spectral slope is different
when measurement are made parallel and perpendicular to the local magnetic 
field. However, in this paper we deal with LOS observations and in this case
it is difficult to measure such anisotropy in intensity maps (see
\citet{Burkhart14}  for a
discussion of how this can be done in velocity centroid maps).
For our purposes in this work we can use the $-5/3$ reference slope
as we deal with LOS observations of super-Alfv\'enic turbulence and not the spectrum as measured in the local frame of reference
to the magnetic field.


In the current paper we are interested in the behavior of the density/column density spectral slope in  compressible self-gravitating turbulence.  
In the presence of supersonic turbulence, such as exists in GMCs, the density spectral slope is shallower than the relations discussed above due to 
shocks creating small scale enhancements of density \citep{Beresnyak05, Kowal07}.
\citet{Burkhart10} plotted the power spectral slope of column density maps
versus sonic Mach number
and found that the slope of the power spectrum of ideal MHD turbulence is increasingly shallow as the Mach number increase. 
 However they found that the power spectral slope begins to saturate
toward -2 (as compared with the 2D slope of  -8/3) for very high sonic Mach number, regardless of Alfv\'enic Mach number.  

Furthermore, the inclusion of gravity in a magnetized turbulent media changes the behavior of the density spectral slope dramatically.  
As gravity further enhances overdense regions,
the spectrum becomes increasingly shallow as more material collects on small scales \citep{Ossenkopf01, Collins12, Federrath13}. 
In some cases, self-gravitating supersonic turbulence produces density structure that  drive the spectral slope toward positive values.
This is in contrast to non-self-gravitating turbulence where the power is dominated by large scale structures and
the power on the smaller scales is decreasing. 

In this section, we investigate the evolution of 
the 1D column density power spectrum (which we denote as $P(k)$) as one transitions from gas dynamics dominated by supersonic MHD turbulence to self-gravitating.  We also propose a model for the power spectrum of a self-gravitating fluid based on our findings.
For the AMR data, the analysis was done on a coarse-grained model, where
the refinement was restricted via volume-weighted average to the coarsest level.
This is due to the sparse sampling of data at these higher wave numbers, which
would lead to unphysical suppression of power at these wave numbers.

 \begin{figure} \begin{center}
\includegraphics[width=\hw\textwidth]{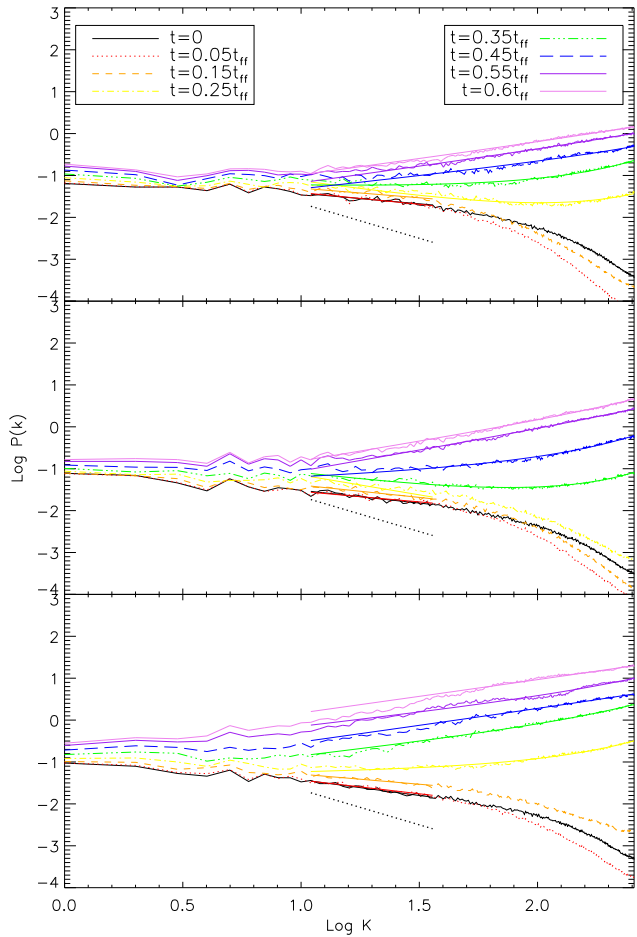}
\caption[ ]{1D Power spectra as a function of wavenumber, averaged over three lines of sight, for the Enzo simulations at resolution $512^3$.
Top to bottom plots represent high to low magnetic field runs, respectively.  Solid colored lines are overplotted curves of Equation \ref{eq:mod} with least square fit
parameters given in Table \ref{tab:fits} for each model. The color scheme is identical to Figure 7.}
\label{psav} \end{center} \end{figure}


We plot the 1D power spectrum averaged over three lines of sight as a function of
wavenumber in Figure \ref{psav} for the Enzo simulations. We overplot a solid
black line in our inertial range with slope of -5/3 from log $k$=1.1 to log $k$=1.4 for reference.
Solid colored lines are overplotted curves of Equation \ref{eq:mod} (next subsection) with least square fit
parameters given in Table \ref{tab:fits} for each model. 
We find that there is no significant variation of the column
density spectra with projection direction, similar to past studies such as
\citet{Federrath13}.

Similar to the PDFs, we find that there exists three distinct stages of evolution in the power spectrum of collapsing column density images:
\begin{enumerate}
 \item  At $t < 0.15t_{ff}$ (henceforth termed ``early"), the cloud is in a purely turbulent regime, and hence the power spectrum exhibits behavior of supersonic turbulence, i.e. negative valued slopes which are shallower than the -5/3 slope.

\item At $0.15 t_{ff} \leq t \leq 0.35t_{ff}$ (henceforth termed ``intermediate"):
As the timestep increase and gravity begins to dominate the small scales (large
$k$), the slope becomes increasingly shallow 
and at the largest timesteps shown, is positive in the inertial range. This is in agreement with past studies of the power spectrum
of collapsing supersonic turbulence \citep{Collins12, Federrath13}. 
The turn over time step from negative to positive occurs around $t\approx 0.25t_{ff}$
but is also dependent on the strength of the magnetic field.
\item At $t > 0.35t_{ff}$ (henceforth termed ``advanced"), gravity dominates the
  power spectrum at large $k$ values and the slopes in both the inertial rage
  and at larger $k$ are positive valued.
\end{enumerate}

It is interesting that the column density power spectrum (which is an observable quantity) exhibits different behavior based on time evolution.
This implies that this tool maybe used on observations to not only determine the properties of turbulence in clouds but also the gravitational
state of cores.
We develop a functional fit
to determine the turnover scale at intermediate times transitioning from turbulence dominated to gravitationally dominated in the next subsection.

\begin{figure} \begin{center}
\includegraphics[width=\hw\textwidth]{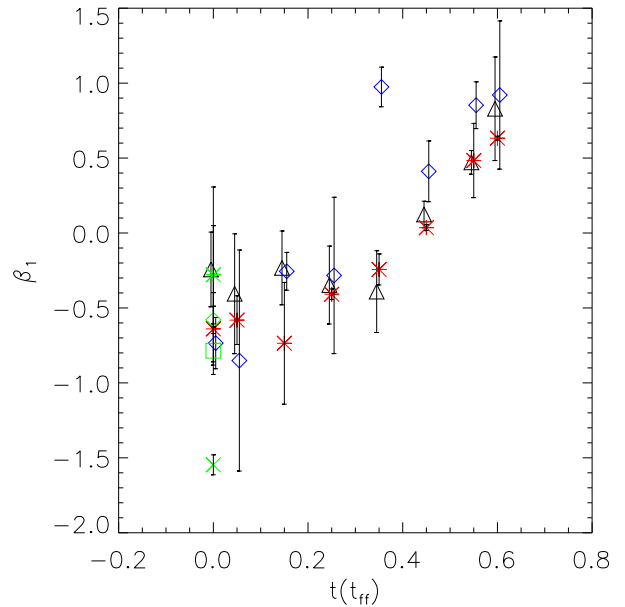}
\caption[ ]{Column density power spectral slopes as a function of time for the Godunov 
and Enzo simulations with varying values of magnetic field. The color-symbols used is similar to Figure 4. Error bars are created from the standard deviation of different sight lines.
The MHD simulations show the expected slopes for subsonic values around $\approx -1.6$ and  becoming increasingly shallow for supersonic turbulence.
The Enzo simulation at t=0 are within the expected slope range of the Godunov/ENO simulations however, once gravity is turned on the  values
of the slope increase past the purely supersonic turbulence cases and eventually become positive.}
\label{fits} \end{center} \end{figure}

We plot the power law slopes (denoted as $\beta_1$) of the spectrum (as shown in
Figure \ref{psav})  versus time for the Enzo and Godunov (green
symbols) simulations in Figure \ref{fits}.
Error bars are calculated by taking parallel and perpendicular sight lines relative to the mean magnetic field.
For each sonic Mach number of the Godunov simulations, two separate Alfv\'enic Mach numbers are plotted.

The Enzo simulations
at $t=0$ show slopes that are fully consistent with the Godunov MHD simulations for the same sonic Mach number.
For $t> 0$ the slopes of the Enzo simulations begin to become increasingly shallow as compared with 
the purely turbulent scenarios, as is evident in Figure \ref{psav}.  However this is not significant
within the error bars until $t > 0.2t_{ff}$. The most significant gains in the value of the slope
are made in the case of the low magnetic field.  At $t >0.25t_{ff}$ this slope becomes positive and remains so 
as the cloud continues to evolve.  For the high and mid magnetic field cases
the column density power spectral slope becomes positive at  $\approx t >0.45t_{ff}$. 
In both this work and previous works such as  \citet{Federrath13},  positive values for the column density
spectral slope are observed at evolved collapse stages.

\subsection{Modeling the Power Spectrum of Self-gravitating Turbulence}
\label{sec.fit}

Figure \ref{psav} shows that the power spectrum of the gas column density map reveals different behavior for
 supersonic MHD turbulence and supersonic MHD turbulence undergoing collapse. Namely that supersonic turbulence displays a negative
power spectral slope while in the collapsing state the power spectrum becomes increasingly shallow past the expectations of 
supersonic turbulence and even becomes positive valued. This has also been
confirmed in other studies, \citet[e.g.][]{Collins12}
and \citet{Federrath13}.

In this subsection we address the turnover scale that occurs as the spectrum transitions from turbulence dominated
to gravitational dominated in order to understand the nature of this transition and to provide observers with additional methods to disentangle 
the dynamics of turbulence from cloud collapse. 

 We propose a functional fit to the power spectrum in the form:

\begin{equation}
 P(k)=A_1k^{\beta_2}exp(-k/k_c)
\label{eq:mod}
\end{equation}
This is characterized by an amplitude $A_1$ and power law behavior $k^{\beta_2}$ which dominates the small $k$ behavior (i.e. turbulence dominated).
The scale $k_c$
characterizes the turn over from turbulence dominated to self-gravity dominated which is observed in the 
power spectrum at an intermediate time evolution (see Figure \ref{psav}). 
The advantage of the exponential form is that it can roughly describe all three stages of evolution discussed in the previous section
and hence could be used by observers who do not know apriori the time scale.
 At $t$=0 ``early'' timesteps the exponential can be negative to describe the
 dissipation of energy (a form used in previous works on MHD turbulence, e.g.
 \citet{Lazarian04}). 
At intermediate times it can describe the turnover scale seen in Figure \ref{psav}. 
Finally at ``advanced" times Equation \ref{eq:mod} reduces again to a power law form as $k_c$ becomes large. 
We fit the function
from $k$=13 to avoid the scales dominated by the turbulence driving.

Figure \ref{psav} overplots the fits of the three parameters ($A_1$, $\beta_2$, and $k_c$ ) given in Equation \ref{eq:mod}
as thick solid lines.
We present the fitted parameters in Table \ref{tab:fits}.
For the power spectra with clear signatures of turbulence, i.e. power law behavior in the inertial range and decreasing
power in the dissipation range (large $k$), we only overplot the contribution of the power law in Equation \ref{eq:mod}.  
We note this in
Table \ref{tab:fits} under the column for $k_c$
with N/A.  Table \ref{tab:fits} also lists the values of the slope $\beta_1$ which is calculated from a linear fit 
in the inertial range from $k$=13 to $k$=24 and plotted in Figure \ref{fits}.

\def\smax{\ensuremath{\langle\Sigma_{\rm{max}}\rangle}}
\begin{table*}
\begin{center}
\caption{$512^3$ Enzo power spectra fit parameters. $B_{ext}$ denotes the
magnetic field regime (see Table 1); t denotes the time step in units of the
free fall time;
$\beta_1$ is the power spectral slope fitting from a standard power law fit;
Columns 4-6 ($A_1, \beta_2, k_c$) show the fit parameters to Equation
\ref{eq:mod}; $L_c$ shows the length scale corresponding to $k_c$
assuming a 4 Mpc box $L_c = 4 \rm{pc} * 2 \pi/k_c$; $\smax$ shows the maximum column density, averaged over all three
axes; $\beta_{t_0+max(t)}$ shows the slope of the turbulence with delta
function, $t_0$+max(t) (Section \ref{sec.SGtoNSG});
$\beta_{\rm{rec}}$ shows the fit to the truncated power spectrum (Section
\ref{NSG_FROM_SG}); $\Sigma_c$ shows the cutoff column density in code units. We can convert the code densities
to physical units by multiplicative scaling factor of 1.4x10$^{22}$ cm$^{-2}$ assuming a cloud of mean density 1000 cm$^{-3}$ and 
size of 4.6 pc.}

\label{tab:fits}
\begin{tabular}{ccccccccccc}
\hline\hline
$B_{ext}$ & t &     $\beta_1$       &$A_1$& $\beta_2$      &$k_c$  & $L_c$ &
\smax  & $\beta_{t_0+max(t)}$ & $\beta_{\rm{rec}}$ &$\Sigma_{\rm{c}}$
\\
\tableline
High &$0t_{ff}$    &-0.34$\pm0.13$  &0.12 &-0.5$\pm0.03$   &  N/A  & N/A   &4.6     &-0.5   $\pm0.03$  &-0.6  &4.40826 \\
     &$0.05t_{ff}$ &-0.42$\pm0.08$  &0.14 & -0.55$\pm0.1$  & N/A   & N/A   &4.8     &-0.54  $\pm0.0$   &-0.8  &4.35521 \\
     &$0.15t_{ff}$ &-0.38$\pm0.06$  &0.14 & -0.45$\pm0.12$ & N/A   & N/A   &9.9     &-0.53  $\pm0.01$  &-0.5  &5.57656\\
     &$0.25t_{ff}$ &-0.31$\pm0.06$  &0.49 &-0.85$\pm0.15$  & 120   & 0.241 &65.7    & 0.39  $\pm0.04$  &-0.4  &5.41682\\
     &$0.35t_{ff}$ &-0.36$\pm0.01$  &0.07 & -0.12$\pm0.05$ & 142   & 0.204 &139.7   & 0.78  $\pm0.02$  &-0.8  &5.24604\\
     &$0.45t_{ff}$ &0.1$\pm0.11$    &0.01 &0.6$\pm0.04$    & 700   & 0.041 &176     & 0.82  $\pm0.01$  &-0.2  &5.46864\\
     &$0.55t_{ff}$ &0.31$\pm0.07$   &0.01 &0.78$\pm0.07$   & 1250  & 0.023 &261     & 0.86  $\pm0.0134$&-0.1  &5.75271 \\
     &$0.60t_{ff}$ &0.41$\pm0.07$   &0.02 &0.76$\pm0.06$   & 1452  & 0.020 &290     & 0.87  $\pm 0.0$  &-0.2  &5.60185 \\
\hline \hline                                                                                                          
Mid &$0t_{ff}$     &-0.7$\pm 0.05$  &0.12 &-0.61$\pm0.03$  &  N/A  & N/A   &4.8     &-0.61  $\pm0.03$  & -1.0 &4.58494 \\
     &$0.05t_{ff}$ &-0.61$\pm 0.11$ &0.09 &-0.5$\pm0.05$   &  N/A  & N/A   &4.6     &-0.61  $\pm0.0$   & -0.6 &4.10902 \\
     &$0.15t_{ff}$ &-0.61$\pm 0.04$ &0.16 &-0.59$\pm0.08$  &  N/A  & N/A   &7.4     &-0.6   $\pm0.0$   & -0.5 &6.79376 \\
     &$0.25t_{ff}$ &-0.46$\pm0.07$  &0.62 &-0.95$\pm0.03$  &  N/A  & N/A   &13.6    &-0.75  $\pm0.03$  & -0.9 &10.1390 \\
     &$0.35t_{ff}$ &-0.27$\pm0.0$   &0.38 &-0.71$\pm0.04$  &  106  & 0.273 &80.3    & 0.63  $\pm0.03$  & -1.1 &5.54317 \\
     &$0.45t_{ff}$ &0.01$\pm0.16$   &0.03 &0.26$\pm0.09 $  &  172  & 0.168 &156     & 0.82  $\pm0.01$  & -0.5 &4.83240 \\
     &$0.55t_{ff}$ &0.5$\pm0.0$     &0.01 &0.98$\pm0.17$   &  1051 & 0.028 &434     & 0.88  $\pm0.0$   & 0.3  &4.60489  \\
     &$0.60t_{ff}$ &0.68$\pm 0.09$  &0.02 &0.91$\pm0.33$   &   565 & 0.051 &479     & 0.89  $\pm0.0$   & 0.1  &4.05968 \\
\hline\hline                                                                                                           
Low &$0t_{ff}$     &-0.8$\pm0.01$   &0.21 &-0.75$\pm0.14$  &  N/A  & N/A   &5.4     &-0.75  $\pm0.14$  &-0.8  &4.56127 \\
     &$0.05t_{ff}$ &-0.77$\pm0.1$   &0.17 &-0.65$\pm0.06 $ &  N/A  & N/A   &7.1     &-0.8   $\pm0.0$   &-1.0  &5.28883 \\
     &$0.15t_{ff}$ &-0.38$\pm0.04$  &0.15 &-0.48$\pm0.12$  &  N/A  & N/A   &30      &-0.75  $\pm0.03$  &-0.6  &6.73403 \\
     &$0.25t_{ff}$ &-0.16$\pm0.02$  &0.04 &0.1$\pm0.13$    &   176 & 0.164 &206     & 0.85  $\pm0.01$  &-0.9  &5.46192  \\
     &$0.35t_{ff}$ &0.78$\pm0.1$    &0.03 &0.65$\pm0.48$   &   350 & 0.083 &655     & 0.89  $\pm0.0 $  &-0.0  &4.45580  \\
     &$0.45t_{ff}$ & 0.64$\pm0.1$   &0.05 &0.79$\pm0.2$    &   2945& 0.010 &784     & 0.89  $\pm0.0 $  &-0.7  &5.62422   \\
     &$0.55t_{ff}$ &0.99$\pm0.09$   &0.16 &0.64$\pm0.18$   &   441 & 0.066 &1175    & 0.89  $\pm0.0$   &-0.7  &12.7582  \\
     &$0.60t_{ff}$ &0.91$\pm0.03$   &0.24 &0.79$\pm0.19$   &   5434& 0.005 &1939    & 0.89  $\pm0.0$   &-0.5  &8.81122   \\

\hline\hline
\end{tabular}

\end{center}
\end{table*}

Figure \ref{psav} shows that  Equation \ref{eq:mod} 
reasonably models the turn over  scale at intermediate time steps and the gravitationally dominated high $k$ scales (i.e. small spatial scales).
The fits are more robust for the high and mid magnetic field cases, whereas the low
magnetic field simulation (bottom panel) is not as well fit for the mid $k$ ranges and  hence for this simulation
the values of $k_c$ are more erratic. 
In general, the $\beta_2$ values from the fit of Equation \ref{eq:mod} listed in
Table \ref{tab:fits} agree
within the error bar with the pure power law fit (i.e. $\beta_1$) for early timesteps.

Table \ref{tab:fits} shows that the turnover scale $k_c$ generally increases with increasing timestep.
This is because at $t >0.5t_{ff}$ the spectrum at the large $k$ scales has completely transitioned from having a
negative slope (turbulence dominated) to having a positive slope (gravity dominated). 
As $k_c$ increases the exponential term influences Equation \ref{eq:mod} less
until again only the power law term dominates again. At ``advanced" time steps the values of $k_c$ are 
equivalent to infinity since they extend beyond the range of the power spectrum being plotted (i.e. past $k$=256).
Thus the $k_c$ scale of interest occurs at intermediate time steps where gravity is just starting
to dominate the power spectrum. 
This is $k_c\approx100-200$ and occurs at timesteps from $t=0.25t_{\rm{ff}}$ to $t=0.45t_{\rm{ff}}$, with the difference being
attributed to the influence of the magnetic field.  Using the scaling for these simulations as reported in 
\citet{Collins12}, this corresponds to a length scale of $L = 0.005-0.2 \rm{pc}$.
Attention to Figure \ref{fig:pdfs} shows that these timesteps are also where the power law tails
begin to form in the PDFs.
Visual inspection of the column density of these snapshots reveals
that this is the moment when the first cores begin to form from the turbulent cloud.
Thus, the turnover scale $k_c$ can be used as an additional diagnostic for 
star formation evolution in a similar manner to the PDFs and the power spectral slope.

 \begin{figure} \begin{center}
\includegraphics[scale=.5]{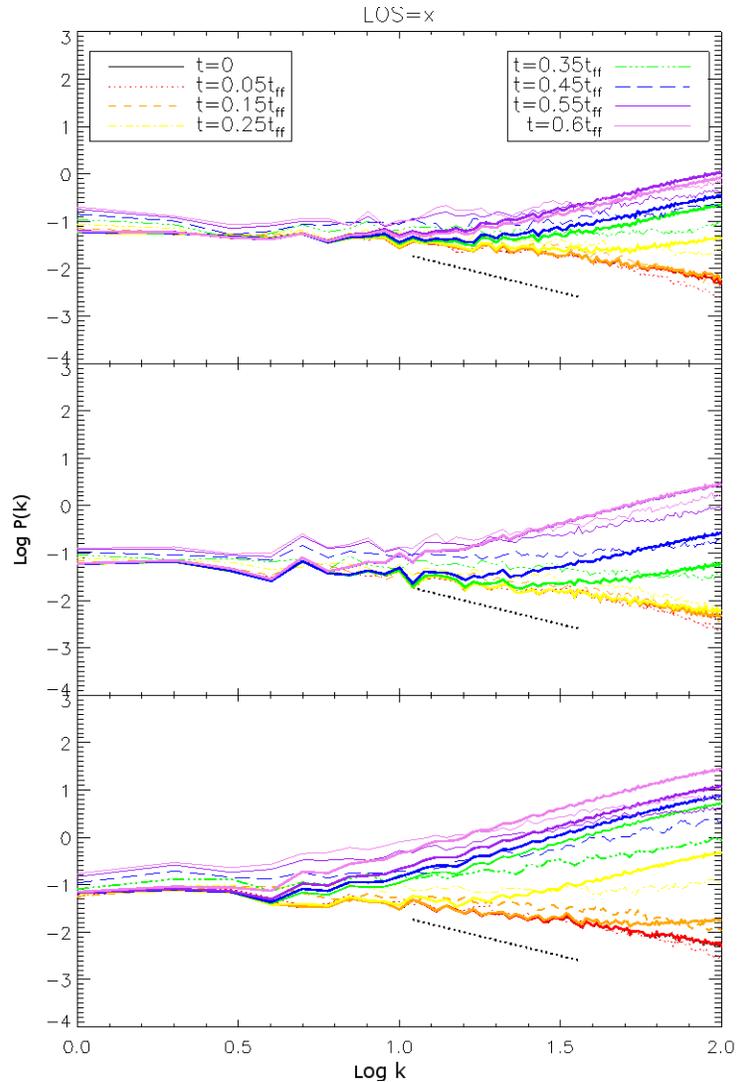}
 \caption[ ]{Reconstructing self-gravitating spectra from turbulent spectra: Power spectrum as a function of wavenumber, averaged over three lines of sight, for the Enzo simulations at resolution $512^3$.
 We overplot (thick solid lines) the power spectrum of the Enzo simulation with
 the $t_0+\rm{max}(t)$ column density map, i.e. the $t$=0 map 
 that has one pixel modified such that it has amplitude equal to the peak value of the given snapshot at $t>0$.}
 \label{pspsf} \end{center} \end{figure}

\subsection{Mimicking Self-gravity in the Turbulence Power Spectrum}
\label{sec.SGtoNSG}

The timestep of $t\approx 0.25t_{ff}$, where gravity begin to dominate the power spectrum and PDFs 
of turbulent magnetized clouds, is the moment when very high density cores begin to form in our simulations.
These cores exist on the smallest scales (i.e. a few pixels), yet influence the global turbulence statistics of the cloud.
Thus in order to understand the affects of the small scales on the large scales a natural question arrises:
can the global properties of gravity be replicated with a 
single (or multiple) point function which replicates a core?

We test if it is possible to replicate the power spectrum of gravoturbulence using the $t$=0 snapshot of the column density (i.e.
the timestep which has no gravity) and the maximum pixel value of the snapshot at some later timestep. 
We will call this map the
``$t_0+\rm{max}(t)$" column density map as it is the original $t$=0 map with one pixel added to it that represents the largest value 
of a snapshot at a later time.

Figure \ref{pspsf} plots the power spectrum similar to Figure \ref{psav} 
but overplots the power spectrum of the t$_0$+max(t) column density map.
In general, the correspondence is fairly good for the 
both the later timesteps and and earlier time steps but has trouble matching
the intermediate timesteps where the turnover from turbulence dominated spectra to gravity dominated spectra takes place.
At these time steps the power spectrum has a more complex behavior, i.e. can not be fit with a simple power law,
which is evident by the lower value of $k_c$ in Equation \ref{eq:mod}.

We fit the t$_0$+max(t) column density map using Equation \ref{eq:mod} and 
report the slopes (denoted as $\beta_{\delta}$)
along with the values of the maximum of the column density map in Table
\ref{tab:fits}.
As is evident from visual inspection of Figure \ref{pspsf}, the largest discrepancies between the actual column density map and the t$_0$+max(t) column density map 
are at the intermediate time steps (t=0.25t$_{ff}$-0.35t$_{ff}$).  Additionally
at these time-steps the maximum value of the column density map (Table
\ref{tab:fits}, column 7)
increases dramatically, signaling that gravitational collapse is 
beginning to dominate the low k values of the power spectra of the gas.
We note that our results would be similar if we had used several delta function-like points or if we had placed them in different areas of the map.

The fact that one can alter the turbulence power spectrum to mimic the
gravoturbulence power spectrum provides researchers with an avenue of comparing
the power spectrum of turbulence simulations with self-gravitating clouds
following the prescription of maximum values given in Table \ref{tab:fits}.   It also indicates
that, on the scales of GMCs, the formation of cores introduce $\delta$-function like 
intensity profiles to the power spectrum.
Finally, if it is possible to mimic the effects of the gravoturbulence power spectrum then it should be feasible to remove the signatures of gravity as well.  We discuss this in the next subsection.

\subsection{Restoring the Turbulence Power Spectrum in Self-gravitating Clouds}
\label{NSG_FROM_SG}

In Section \ref{sec.SGtoNSG} we demonstrated that it is possible to reproduce the
self-gravitating power spectrum from the turbulent power spectrum, 
 by adding a suitably chosen $\delta$ function
to the turbulent spectrum. This suggests that the removal of a $\delta$ function
from the self-gravitating spectrum can recover the turbulent spectrum.  
It was first shown by \citet{Beresnyak05} that, for supersonic turbulence, the power
the spectrum of $\rho$ is made steeper by reducing rare density peaks, either by
restriction or the logarithm.   Here we try to use the relatively flat column
density spectrum to recover the steeper turbulent by a suitably chosen
restriction.

\begin{figure*}  \begin{center}
\includegraphics[scale=.2]{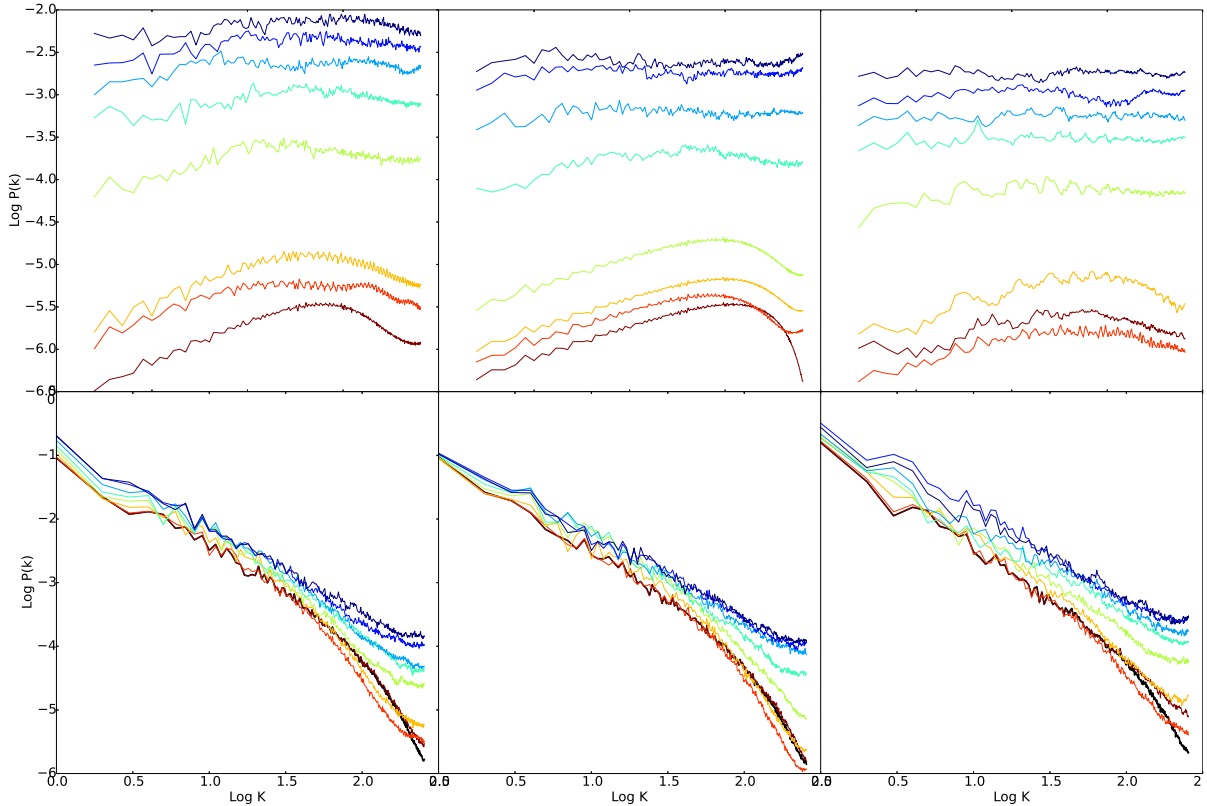}
  \caption[ ]{Top row: ``upper'' spectra,
  $P_{>c}(k)$ which has been constructed to be flat.  Bottom row: ``truncated''
  spectra, $P_{<c}(k)$.  Most of the evolution of the cloud due to the
  collapsing gas has been removed in this bottom row.  Left column shows the
  high field run, while the right column shows the low field run.}
\label{turb_spectra_recovery}   \end{center} \end{figure*}

We define
\begin{align}
  \Sigma_{<c} = \begin{cases} \Sigma& \Sigma < \Sigma_c \\
    0 &\Sigma > \Sigma_c
  \end{cases}  \\
  \Sigma_{>c} = \begin{cases} 0& \Sigma < \Sigma_c \\
    \Sigma &\Sigma > \Sigma_c
  \end{cases} 
\end{align}
to be truncated column densities, and $P_{<c}(k),\ P_{>c}(k)$ be the power spectra of
each.  We saw in Section \ref{sec.SGtoNSG} that one could construct a new field,
\def\sigmansg{\Sigma_{\rm{NSG}}}
\def\sigmamax{\Sigma_{\rm{max}}}
\begin{align}
\Sigma_1 = \sigmansg(\bf{x}) + \sigmamax(t) \delta(\bf{x})
\end{align}
where $\sigmansg$ is the column density before the action of self gravity, and
$\sigmamax(t)$ is the peak column density at some time $t$, then the
power spectrum of $\Sigma_1$ is quite similar to the power spectrum of the self
gravitating cloud at time $t$.  Since it can be shown that the power spectrum of
$\delta(\bf{x})$ is a constant with respect to $k$, we aim to select a cutoff
column density, $\Sigma_c$, that has a power spectrum that is similarly constant
over a given range.   Thus for some self-gravitating cloud, if $P_{>c}(k) = const$
for some threshold $\Sigma_c$, then we show that the spectrum of the lower
column density gas, $P_{<c}$ recovers the initial turbulent spectrum reasonably
well.  We show this in Figure \ref{turb_spectra_recovery}.  In this figure, the
left column is the high magnetic field simulation, and the right is the low magnetic field
simulation.  The top row  shows $P_{>c}$,
which $c$ is chosen to produce flat spectra; the bottom row shows $P_{<c}$,
the truncated spectra.  
The black line is $P_{\sigmansg}$, the initial turbulent state, and
the colored lines are spectra for subsequent collapse states, as in earlier
figures.  To ensure $P_{>c}$ is flat we used a recursive bisection
technique, varying $\Sigma_c$ until the slope of $P_{>c}$ for $k>10$ was near zero. 
  While there is significant evolution in the full column
density power spectrum (e.g. Figure \ref{psav}), the evolution of the slope of the truncated
spectrum is significantly restricted.  While this technique is promising, it is
is not yet able to discriminate between the initial turbulent conditions. The
slope of the full spectrum, $P(k)$, for the turbulent state in the high field
run, for $\log k$ between 1.1 and 1.4, is $-0.5$, while the truncated slopes
vary from $-0.8$ to $-0.2$.  For the mid field, the full slope is $-1.0$ and the
variation is from $-1.0$ to $0.1$.  For the low field, the full slope is
$-0.8$, and the truncated slope varies from $-0.9$ to $0.0$.  The last column
of Table
\ref{tab:fits} shows the fits.  We hope that with
some refinement this technique will allow for the further separation of
collapsing and turbulent clouds.

\section{Discussion}
\label{sec.discussion}

Turbulence in GMCs is believed to be a part of a cascade that extends over 12
orders of magnitude in scale \citep{Armstrong95}. 
In this paper we study the observable signatures of self-gravitating magnetized supersonic turbulence by applying the probability density functions (PDFs) and the spatial density power spectrum to synthetic column density maps
generated with
the Enzo code. Unlike other recent studies (e.g. \citet{Federrath13})  that relate  density diagnostics to star formation efficiencies/rates that rely on sink particle prescriptions, we
stick entirely to observable column density statistics and do not use any numerical prescriptions for star formation.  In this sense we study the effects
of gravitational collapse on driven turbulence and the evolution of the cloud as collapse proceeds.

We find that there exists three characterizable stages of the evolution of the PDFs and density power spectrum
of the collapsing cloud which we term "early," "intermediate," and "advanced."
The natural question that arises is how well do the predictions made in this paper match up with observations?
The PDF of molecular clouds has been studied in a number of recent surveys and we have over-plotted measurements of the
powerlaw tail index of those published in \citet{Schneider13,Schneider14}  Figure  \ref{fig:alpha2}.  
These clouds include 
NGC2603, which is a high-mass active star forming regions,  the Auriga cloud,
which is a low-mass star formation region, and the Orion B and Aquila clouds, which are moderately star forming
\citep{Schneider13b}.  Additionally, \citet{Schneider13b} published the
Herschel data for the low-mass star forming cloud called Maddalena, which
they found has a power law slope values of -3.65, however we do not include this
value in  Figure \ref{fig:alpha2} as it is steeper than the range plotted. 

How well do the ages and star formation history match with trends predicted by the simulations in Figure  \ref{fig:alpha2}?
We discuss some of the literature on each of these clouds in order of evolutionary age predicted by the PDF in Figure  \ref{fig:alpha2} (youngest to oldest).
\begin{enumerate}
 \item The Maddalena cloud has 41 young stars with disks and 33 protostars in the center of the cloud with an age estimate of a few Myr
\citep{Megeath09}.

\item Aquila's age estimates are expected to be a few Myr \citep{prato08}.
\item  Auriga-California age estimates are not well constrained but it is suggested this cloud is not very evolved based on a high fraction of Class I and Class F YSOs \citep{Broekhoven-Fiene14}.
  \citet{Harvey13} tabulated 60 likely pre-main-sequence objects while \citep{Lada10} report 175-279 YSOs. 
\item Orion B  has approximately 635 YSOs \citep{Lada10}.
 
\item  The age estimates of NGC3603 are $10-20$Myrs \citep{Beccari10} making it the oldest cloud of the ones we compare with here.
 There are more than 10,000 stars with 7000 young stars (forming a single power
law IMF) in NGC3603 \citep{Harayama08}.
\end{enumerate}

In light of the literature on these clouds it would seem that the PDFs  are able to trace the  evolutionary state of the cloud.
This bodes well for future observational studies, which should combine the measures presented in this paper including the PDF variance, skewness, kurtosis, Tsallis and power spectrum 
to dissect the evolutionary state of clouds.


Our results also revisit the now well established debate over the impact of
magnetic fields on the star formation process.  The idea that magnetic fields
wholly dominate the formation of stars is accepted with less enthusiasm these
days.  Indeed, within the dynamical picture of the interstellar medium there are
factors in addition to the magnetic field that impede the gravitational collapse
of a molecular cloud and make star formation less efficient, comparable to the
observed rates \citep[see][]{McKee07}.  For instance, turbulent magnetic fields
undergoing fast reconnection  \citep{Lazarian99}, which results in reconnection
diffusion \citep{Lazarian12}, diffuse  out of clouds orders of magnitude faster than the typical
ambipolar diffusion timescale that is usually invoked in the traditional theory
of molecular cloud evolution.  Our results here show that it is wrong to
disregard the role of magnetic fields completely, as we see important
dependencies in the evolution of the density statistics on the magnetization of
the media.  Furthermore we note that Table 2 shows values of our dense core regions
being within the range predicted for gravitational free fall to become faster than the reconnection
diffusion timescale  (i.e. around $10^{23} cm^{-2}$, see \citet{Lazarian12}.  Future works should
quantify the role of reconnection diffusion in gravo-turbulent simulations such as the ones studied in this work.

 Currently the gravitational state of a molecular cloud is found by
  comparing kinetic and gravitational energy through the virial parameter,
  $\alpha = 5 \sigma^2 R/M$, which requires knowledge of the clouds linewidth,
  $\sigma$, size, $R$, and mass, $M$, as well as the implicit assumption that
  $\sigma$ and $M$ are spatially correlated.  The techniques presented here
  provide a potential tool to probe the evolutionary state of a cloud using
  only the column density, which eliminates several sources of uncertainty in
  the estimation and implicitly includes information about magnetic field
  strength.  Future numerical studies should focus on varying the sonic and Alfv\'en Mach numbers
  in order to change both the power spectral slope and dynamical importance of the magnetic field
  in order to determine how turbulence speeds up or impedes star formation.  The filtering techniques
  discussed in this paper can illuminate the dynamics of turbulence even if the cloud is star forming and 
  in an advanced state of 
  collapse.

The current simulations aim to isolate the effects of magnetic fields and
  gravity on supersonic turbulence.  However the use of periodic boundary
  conditions, solenoidal driving,  and the lack of feedback are potential sources of discrepancy with
  real star forming clouds.   Real clouds form from some sort of compressive
  flow, be it cloud collisions, gravitational instability, or thermal
  instability \citep[and references therein]{Dobbs13} and this has been shown to
  impact power spectra in such simulations \citep{Federrath13}.  However it is
  possible that in the inertial range the solenoidal-to-compressive ratio
  reaches a universal value \citep{Kritsuk10b}.  Further the periodic boundary
  condition imposes a conservation of total volume that may artificially enhance
  the amount of low density gas in the cloud.  The impact of feedback is in
  general to inject kinetic energy, which steepens the slope by suppressing
  small scale structure \citep{Sun06}.   Future simulations will incorporate
  these effects.

Although the literature is abounding with papers regarding the 3D density PDF and its relation to the star formation efficiency, 
 the statistical properties of velocity and
magnetic field are also of vital importance to describe and understand
the turbulence cascade and the star formation process.  Further, these statistical properties are essential to
differentiate between the "log jam" of theoretical models of star formation.
Much work has gone into developing a toolbox of measurements to ascertain the
physical conditions in GMCs and the ISM.  The work presented here is a 
step towards extending this toolbox to self-gravitating turbulence.   
Techniques such as the Velocity Coordinate Spectrum (VCS, \citet{Lazarian09}) can provide the injection scale of turbulence and
turbulence energy density and compare the observed spectrum of clouds to analytical predictions.
The studies
presented here can be improved upon in the future using techniques that go beyond the PDFs and the power spectrum; for instance, earlier research with non-self-gravitating turbulence has
shown that the anisotropy of velocity fluctuations can be used to find the
magnetization of the medium \citep{Lazarian01, Esquivel05, Esquivel10,
Burkhart14}, especially if used in conjunction tools to measure
the sonic Mach number \citep[see][and references therin]{Burkhart13}. 

\def\Ms{\ensuremath{{\cal M}_s}}
\def\Ma{\ensuremath{{\cal M}_A}}
It is worth summarizing the impact of the three parameters explored here (Mach
number \Ms, Alfv\'en Mach number \Ma, and collapse state) on each
of the diagnostics available to us:
\begin{enumerate}
\item \emph{Column Density PDF power law tails} depend strongly on collapse state and \Ma.
  \item \emph{Column Density variance from log-normal fit} depends strongly on
    \Ms, slightly on \Ma, and little on collapse state.
 \item \emph{Column Density variance from direct measurement of column density}
    is sensitive to the collapse state and  \Ms, and moderately correlated
    with \Ma.
  \item \emph{Skewness and Kurtosis of $\zeta$} are not strongly correlated with \Ms; they
    are not strongly correlated with \Ma, but are strongly correlated with
    collapse state.
  \item \emph{Tsallis fit parameters: q, w, A} depend strongly on collapse state and \Ma.
  \item \emph{Density Power Spectrum} is a very strong indicator of collapse state.
  \item \emph{Velocity Power Spectrum}: While not studied in this work and
    included here for completeness, it was shown by \citep{Collins12} and
    \citep{Federrath12} that the velocity power spectrum is sensitive to \Ms\
    and \Ma, but not the collapse state. We believe that the observed low correlation of the velocity
power spectrum with the collapse state is the consequence of the relatively
low energy injection to the turbulence cascade during the collapse as compared to the 
driving energy.  This matter should be explored further in the future in simulations
and in observations using the VCS/VCA methods \citep{Lazarian09}.
\end{enumerate}

The differential sensitivity of the column density statistics on the key parameters of 
self-gravitating turbulence opens up prospects for determining the sonic Mach number,
Alfv\'enic Mach number and the collapse state by combining different statistics. 

\section{Conclusions}
\label{sec.conclusions}

Turbulence, magnetic fields, and gravity are some of the key ingredients in the star formation process.
Using synthetic column density maps
generated with 
the Enzo AMR code, we
investigated the robustness of the probability density functions and the spatial power spectrum
for understanding and separating the roles of magnetic fields,  supersonic motions and gravitational collapse
in the observable column density distribution.
The PDFs and power spectrum reveal three stages of cloud evolution as it progresses from diffuse turbulence dominated to collapse
dominated.

Regarding the PDFs, the statistical moments and the Tsallis fit to the observable column density distribution we find that:

\begin{enumerate}
\item For early times, i.e. $t<0.15t_{ff}$,  the cloud has a lognormal distribution and will develop a high density power law tail at intermediate
times between  $t$=$0.25t_{ff}-0.35t_{ff}$.  The development of the tail, including its
slope, also depends on the magnetization of the cloud.  The PDF power law tail forms earlier in 
clouds with lower magnetization. The tails then become increasingly shallow as the collapse proceeds with time.
\item The directly calculated variance of the column density map (and natural logarithm of the column density map),
 is a sensitive diagnostic for the cloud evolution. The variance increases monotonically with time as the cloud collapses and  depends on the magnetic field strength. This
increase in variance with collapse causes the cloud to deviate from the 
sonic Mach number - variance relationship expected from non-self-gravitating
turbulence.

\item The skewness and kurtosis of the  natural logarithm of the column density map are insensitive to the sonic Mach number
and trace the collapse of the cloud with time, i.e. they track the formation of the power law tail.
\item The three Tsallis fit parameters for the incremental PDFs all strongly trace the evolution of the collapse with time: higher values
correspond to more evolved cloud collapse.
A strong magnetic field in a gravoturbulent cloud produces lower values of the amplitude and width of the incremental PDFs.
\end{enumerate}

We find the spatial power spectrum to be complimentary to the PDFs for studies of the evolutionary state of collapsing clouds.  In particular we find that:
\begin{enumerate}
\item  The column density power spectrum of supersonic self-gravitating turbulence shows characteristics of a turbulence only power spectral slope
at early stages of collapse ($\approx 0.15t_{ff}$).  At intermediate time-steps
$t$=$0.25t_{ff}-0.35t_{ff}$, the inertial range slope becomes increasingly shallow and the dissipation range 
curves upwards.  Eventually at advanced times the slope becomes positive and a positive sloped power law is seen down to the dissipation scales.  
The timescales of the changes in the slope depend on the magnetic field, with lower magnetic field
facilitating earlier increases in the power spectral slope.
\item We fit a three parameter function to the gravoturbulence power spectrum:  $P(k)=A_1k^{\beta_2}e^{-k/k_c}$, where $A_1$ describes the amplitude,
 $k^{\beta_2}$ describes the power law behavior and the scale $k_c$
characterizes the intermediate stage turn over from turbulence dominated to self-gravity dominated which is observed in the 
power spectrum at times between $\approx0.25t_{ff}-0.35t_{ff}$ depending on the strength of the magnetic field.
The exponential term only becomes important (i.e. $k_c$ becomes on the order of 100) when the first cores begin to form in our simulations.  
The power law slope of the fit, $\beta_2$ is comparable to the traditional linear fit power law slope in the inertial range. 
\item We find that the effects of self-gravity on the power spectrum can be mimicked in a turbulence only simulation 
by including a single point in the column density map that is equivalent to the maximum valued point in a self-gravitating map at a given 
time-step.  This addition of a point source is reminiscent of including delta-function like behavior to the power spectrum at high k values.
We provide values of the maximum points at each given time step and find that the slopes reasonably match the 
time-steps when $k_c$ is very large and the exponential term to the gravoturbulence power spectra relation is negligible. 
\item We find that the effects of self-gravity on the power spectrum can be removed and the turbulence power spectrum restored
through spatial filtering of the high density material.  
\end{enumerate}

\acknowledgments 
The authors would like to thank the referee for constructive comments and suggestions which greatly improved this paper.
B.~B. is grateful for support from the NASA Einstein Fellowship.
A.~L. is supported by the NSF grant AST 1212096.
B.~B. and A.~L. are grateful for support from the Center for Magnetic Self-organization in Labortory and Astrophysical Plasmas (CMSO).
Computer time was provided through NSF TRAC allocations TG-AST090110
and TG- MCA07S014 and the XSEDE allocation TG-AST140008. 
The computations were performed on Nautilus and Kraken at the National Institute
for Computational Sciences (http://www.nics.tennessee.edu/) and on Stampede
and Maverick at the Texas Advanced Computing Center (https://www.tacc.utexas.edu)

\bibliographystyle{apj}
\bibliography{apj-jour,ms.bib}  

\end{document}